\documentstyle[preprint,aps,psfig]{revtex}
\begin{document}
 
\title{Introducing Protein Folding Using Simple Models}
 
\author{D. Thirumalai and D. K. Klimov}
 
\address{Department of Chemistry and Biochemistry and
Institute for Physical Science and Technology\\
University of Maryland, College Park, MD 20742}

\maketitle
\tightenlines

\vspace{0.5in}

\textbf{Key Words: Biomolecular Folding, Symmetry, Designability,
Evolution}


\section{INTRODUCTION}

Most reactions in
cells are
carried out by enzymes \cite{Stryerbook}. In many instances the
rates of enzyme-catalyzed reactions are enhanced by million fold. A
significantly large fraction of all known enzymes are proteins which are
made from twenty naturally occurring amino acids. The amino acids are
linked by peptide bonds to form polypeptide chains. The primary sequence
of a protein specifies the linear order in which the amino acids are
linked. To carry out the catalytic activity the linear sequence has to
fold to a well defined three dimensional structure. In cells only a 
relatively small 
fraction of proteins require assistance from chaperones (helper
proteins) \cite{Lorimer96}. Even in the complicated cellular environment
most proteins fold
spontaneously upon synthesis. The determination of the three
dimensional folded structure from the one dimensional primary sequence 
is the most popular protein folding problem.

For a number of years the protein folding problem remained only of
academic interest. The synthesis of proteins in cells was described by
Crick \cite{Stryerbook}. Schematically this process can be represented as
$DNA \rightarrow RNA \rightarrow Proteins $.
This proposal and
Anfinsen's \cite{Anfinsen73} demonstration that a denatured protein would
fold to
the native conformation under suitable conditions was sufficient to
understand the role of protein
folding in cells. However, in the last couple of decades many diseases
have been directly linked to protein folding (especially
misfolding) \cite{Lansbury99}. Thus,
there is an urgency to understand the mechanisms in the formation of
folded structures. Biotechnology industry is also interested in the
problem because of the hope that by understanding the way polypeptide
chains fold one can design molecules (using natural or synthetic
constituents) of use in medicine. Finally, the full potential of the human
genome project involves understanding what the genes encode. For all these
profoundly important reasons the protein folding problem has taken center
stage in molecular biology. 

Because this problem is complex several venues of attack have been devised
in the last fifteen years. A combination of experimental developments
(protein engineering, advances in X-ray and NMR, various time resolved
spectroscopies, single molecule manipulation methods) and theoretical
approaches (use of statistical mechanics, computational strategies, use of
simple models) \cite{Onuchic97,Dill97,Thirumalai99} has lead to a greater
understanding of how polypeptide
chains reach the native conformation. 

From our perspective there are four
major problems that comprise the protein folding enterprise. They are:

a) Prediction of the three dimensional fold of a protein given only the
amino acid sequence. This is referred to as the structure prediction
problem. In Fig. (1) we show the three dimensional structure of
myoglobin. The structure prediction problem involves determining this fold
using the primary sequence which is given in the lower part of Fig. (1).

b) Given that a sequence folds to a known native structure what are the
mechanisms in the transition from the unfolded conformation to the folded
state? This is the kinetics problem whose solution requires elucidation of
the pathways and transition states in the folding process.

c) How to design sequences that adopt a specified fold
\cite{ShakhFD98}? This is the
inverse
protein folding problem that is vital to the biotechnology industry.

d) There are some proteins that do not spontaneously reach the native
conformation. In the cells these proteins fold with the assistance of
helper molecules referred to as chaperonins. The
chaperonin-mediated
folding problem involves understanding the interactions between proteins.

It is not the aim of this chapter to introduce the reader to all the areas
listed above. Our goal is modest. We describe some of the theoretical
developments which arose from studies of caricatures of proteins. 
Such models were designed to understand
certain
general features about protein structures and how these are kinetically
reached. To keep
the bibliography compact we mostly cite review articles. The interested
reader can find the original papers in these cited works. We hope that
this short introduction will entice the reader to delve  into the ever
surprising world of biological macromolecules.

\section{RANDOM HETEROPOLYMER AS A CARICATURE OF PROTEINS}

In homopolymers all the
constituents (monomers) are identical, and hence the
interactions between the monomers and between the monomers and the solvent
have the
same functional form. To describe the shapes of a homopolymer (in the 
limit of large molecular weight) it is sufficient to model the chain
as a sequence of connected beads. Such a model can be used to
describe the shapes that a chain can adopt in various solvent
conditions. A measure of shape is the dimension of the chain as a
function of the degree of polymerization, $N $. If $N $ is large then 
the precise chemical details do not affect the way the size scales
with $N $ \cite{degennesbook}. In such a description a homopolymer is
characterized in
terms of a
single parameter that essentially characterizes the effective interaction
between the
beads which is obtained by integrating over the solvent
coordinates. 

Proteins are clearly not
homopolymers because many energy scales are required to characterize the
polypeptide chain. Besides the excluded volume interactions and hydrogen
bonds the potential between the side chain depend on the nature of the
residues \cite{Stryerbook}. Therefore, as a caricature of proteins the
heteropolymer model is
a
better approximation. A convenient limit is the random heteropolymer for
which approximate analytic treatments are possible \cite{Garel94}. In a
random
heteropolymer the interactions between the
beads are assumed to be randomly distributed. Some of the interactions are
attractive (which is responsible for conferring globularity to the chain)
while others are repulsive and these residues are better accommodated in an
extended conformation. In proteins water is a good solvent for polar
residues while it is poor solvent for hydrophobic residues. (In a
good solvent contacts between the monomer and solvent are favored
whereas in a poor solvent the monomers are attracted to each
other). Because only
55\% of the residues in proteins 
are hydrophobic it is clear that in a typical protein
energetic frustration plays a role. In addition because of chain
connectivity there is also topological frustration. This arises because
residues that are proximal tend to form structures on short length scales.
Assembly of such short short length scale structures would typically be
incompatible with the global fold giving rise to topological frustration.
Even if energetic frustrations are eliminated a polypeptide chain (in fact
biomolecules) are topologically frustrated \cite{Thirumalai99}.  

In the field of spin glasses and structural glasses such frustration
effects are well known \cite{Mezard87}. Thus, it was natural to suggest
that random
heteropolymers could serve as a simple representation of polypeptide chains.
In Appendix A we sketch computational details for one model of random
heteropolymers. 
Bryngelson and Wolynes proposed, using phenomenological arguments, that
the
Random Energy Model (REM) would be an appropriate description of some
aspects of proteins \cite{Bryngelson87,Onuchic97}. The rationale for this
is
the following: Consider the
exponentially large number of conformations. Because of the presence of
several conflicting energies in a polypeptide chain it is natural to assume
that these energies are randomly distributed. If there are no correlations
between these energies and if the distribution is Gaussian one gets the REM.
Of course, in the REM model the chain connectivity is ignored and there is
no manifest for a spatial dependence of the chain coordinates. We show
in
Appendix B that in the compact phase the  random heteropolymer is
equivalent to REM. 

The random heteropolymer models of proteins are interesting from a
statistical mechanical perspective. However, they do not explain the key
characteristics of proteins such as reversible and cooperative folding
to a unique native conformation. Moreover, 
the theories for heteropolymers suggest that typically the energy
landscapes for these systems are extremely rugged consisting of many minima
that are separated by barriers of varying heights \cite{Garel94}. This
would mean that
kinetically it would be impossible for chain with  a typical realization
of interactions to reach the ground state in finite time scales. Thus the
dynamics of such random heteropolymer models typically exhibits glassy
behavior. Natural proteins do not exhibit any hall mark of glassy dynamics
at most temperatures of interest. It follows that a certain refinement of
the random heteropolymers is required to capture protein-like
properties. One of the important theoretical advances is the
observation that very simple
minimal models \cite{Dill95} can be constructed that capture many (not
all) of the salient features observed in proteins \cite{Dill95}. The
simplest manifestation
of such models are the lattice representation of polypeptide chains. In
the next section we introduce the models and describe a few results that
have been obtained by numerically exploring their behavior.

\section{LATTICE MODELS OF PROTEINS}

The computational protocol for describing protein folding mechanisms is
straightforward in principle. The dynamics are well described by the
classical equations of motion. Simulations of a monomeric protein involves
equilibrating the polypeptide chain in a box of water molecules at the
desired temperature and density. If an appropriately long trajectory is
generated then the dynamics of the protein can be directly monitored. There
are two crucial limitations that prevent a straightforward application of
this approach to study the folding of proteins. First, the interaction
potentials or the force fields for such a complex system are not precisely
known.  Molecular dynamics simulations in the standard packages use
potentials that rely on the transferability hypothesis i.e., interactions
designed in one context can be used in aqueous medium and for larger
systems. 
The need to compute potentials that can
be used reliably in the simulations of protein dynamics remains acute. 

The
second problem is related to
the limitations in generating really long time trajectories that can sample
all the relevant conformational space of proteins. To observe reversible
folding of even a moderate sized protein requires simulations that span the
millisecond time scale. More importantly making comparisons with
experiments involves generating many (greater than perhaps 100) folding
trajectories so that a reliable ensemble average is obtained. Thus, we need
to make progress on both (force fields and enhanced sampling techniques in
long time simulations) fronts before a straightforward all atom simulations
become routine.

In light of the above mentioned difficulties various simplified models of
proteins have been suggested \cite{Dill95}. The main rationale for using
such drastic
simplifications is that a detailed study of such models can enable us to
decipher certain general principles that govern the folding of proteins 
\cite{Onuchic97,Dill97,Thirumalai99}. For
these class of models detailed computations without sacrificing accuracy is
possible. Such an approach has yielded considerable insights into the
mechanisms, time scales, and pathways in the folding of polypeptide chains.
In this section we will outline some of the results that have been obtained
(largely from our group) with the aid of simple lattice models of proteins.

In the simple version of the lattice representation of proteins the
polypeptide chain is modeled as a sequence of connected beads. The beads
are confined to the sites of a suitable lattice. Host of the studies
have used cubic lattice. To satisfy the excluded volume condition only one
bead is allowed to occupy a lattice site. If all the beads are identical we
have a homopolymer model whose characteristics on lattices have been
extensively
studied. To introduce protein-like character the interactions between beads
(ones separated by at least three bonds) that are near neighbors on a lattice
are assumed to depend on the nature of the beads. The energy of a
conformation, specified by \{$r_{i},i=1,2,.....N$\}, is
\begin{equation}
E(\{r_{i}\})=\sum_{i<j}\Delta (\mid r_{i}-r_{j}\mid -a)B_{ij}
\end{equation}
where $N$ is the number beads in the chain, $a$ is the lattice spacing, and $%
B_{ij}$ is the value of the contact interaction between beads $i$ and $j$.
We will consider different forms of $B_{ij}$. Since this model can be viewed
as a coarse grained representation of the $\alpha $-carbons of the
polypeptide chain
the value of $a$ is typically taken to be about 3.8\AA .

Lattice models have been used for a long time in polymer physics
\cite{Orr47}. They were
instrumental in computing many properties (scaling of the size of the
polymer with $N$, distribution of end-to-end distance etc.) of real
homopolymer chains. In the context of proteins lattice models were first
introduced by Go and coworkers \cite{Go75}. The currently popular Go model
considers
only interactions between residues (beads on the lattice) that occur in the
native (ground) state. Thus, in this ''strong specificity limit'' only
native
contacts are taken into account. It follows that in this version of the Go
model the chain is forced to adopt the lowest energy conformation at low
temperatures. Go also considered a variant of this model in which certain
non-native contacts are allowed. Although these models were insightful Go
and coworkers did not use them to obtain plausible general principles of
protein folding. This was partly due to the fact that in their studies they
typically used long chains, and hence exact enumeration was not possible. 

Simple lattice models, with the express purpose of obtaining minimal
representations of polypeptide chains, were first suggested by Chan and
Dill \cite{Chan89}. In order to account for the major interactions in
proteins these
authors argued that the twenty naturally occurring amino acids can be roughly
divided into categories, namely, hydrophobic (H) and polar (P). Chan and
Dill have suggested that this simple HP model can capture many salient
features of proteins. They also suggested that many of the conceptual
puzzles (Levinthal paradox in particular) could be addressed by
systematically studying short chains. This simple exactly enumerable HP
model and their variants  have been used to understand cooperativity,
folding kinetics, and designability of protein structures
\cite{Dill95}. Thus, it is
instructive to describe the calculations that have been done using lattice
models. A study of such models indeed provides a god introduction to the
computational aspects of protein folding.

\textit{Emergence of Structures:} The sequence space of proteins is
extremely dense.
The number of possible protein sequences is $20^{N}$. It is clear that even
by the fastest combinatorial procedure only a very small fraction of such
sequences could have been synthesized. Of course, not all of these
sequences will
encode protein structures which for functional purposes are constrained to
have certain characteristics. A natural question that arises is
how do viable protein structures emerge from the vast sea of sequence space?
The two physical features of folded proteins are : (1) In general native
proteins are compact but not maximally so. (2) The dense interior of
proteins are largely made up of hydrophobic residues and the hydrophilic
residues are better accommodated on the surface. These characteristics
give the folded proteins a lower free energy in comparison to all
other conformations. 

Lattice models are particularly suited for answering the question posed
above. We will show that the two physical restrictions are sufficient to
rationalize the emergence of very limited (believed to be only on the
order of a
thousand or so) protein-like structures. To provide a plausible answer to
this question using lattice models we need to specify the form of the
interaction matrix elements 
$B_{ij}$. For purposes of illustration we consider the random bond (RB)
model in which the elements $B_{ij}$ are distributed as

\begin{equation}
P(B_{ij})=\frac{1}{\sqrt{2\pi }\sigma }\exp [-\frac{(B_{ij}-B_{0})^{2}}{%
2\sigma ^{2}}].
\end{equation}
Here $\sigma (=1)$ is the variance in $B_{ij}$ and the hydrophobicity
parameter $B_{0}$ is the mean value. We chose $B_{0}=0$\ (half the beads are
hydrophobic) and $B_{0}$ $=-0.1$. The latter is motivated by the observation
that in natural proteins roughly 55\% of the residues are hydrophobic
\cite{KlimovProt96}. 

Protein-like structures are not only compact but also have low energy. With
this in mind we have calculated the number of compact structures (CSs) as
CSs with low energy for a given $N$. The number of CSs in its most general
form may be written as
\begin{equation}
C_{N}(CS)\simeq \overline{Z}^{N}Z_{1}^{N\frac{d-1}{d}}N^{\gamma _{c}-1}
\end{equation}
where $\ln \overline{Z}$ is the conformational free energy (in units of $%
k_{B}T$), $Z_{1}$ is the surface fugacity, d is the spatial dimension, and $%
\gamma _{c}$ is measures possible logarithmic corrections to the free
energy. It is clear that natural proteins are relatively unique and hence
their number on an average has to grow at rates that are much smaller than
that given in Eq. (21).To explore this we have calculated by exact
enumeration the number of compact structures, $C_{N}(CS)$ and the number of
minimum energy compact structures $C_{N}(MES)$ as a function of $N$. 

We performed  
exhaustive enumeration of
all self-avoiding conformations, to explore the conformational space
of the polypeptide chain of a given length. In order to reduce the
sixfold symmetry on the cubic
lattice we fixed the direction of the first monomeric bond in all
conformations.  The remaining conformations are related by
eightfold symmetry on the cubic lattice
(excluding the cases when conformations are completely confine to a
plane or straight line). To decrease further the number of
conformations to be analyzed the   Martin algorithm
\cite{Thirumkorea} was modified to
reject all   conformations related by symmetry.

We define MES as those conformations, whose energies lie within the energy
interval \(\Delta\) above 
above the lowest energy \(E_{0}\). Several values for
$\Delta $ were used to ensure that no qualitative changes in the
results are observed. We set \(\Delta\) to
be constant and equal to \(1.2\) (or 0.6) 
(definition (i)). We have also tested another
definition for $\Delta $, according to which  
\(\Delta =1.3|E_0-tB_0|/N\), where $t$ is the number of nearest neighbor
contacts in the ground state (definition (ii)). It is worth noting
that in the latter case  $\Delta $ increases with $N$.  
Both definitions yield equivalent results.
Using these definitions for $\Delta $, we computed $C$(MES) as a function
of the number of residues \(N\).  

The computational technique involves exhaustive enumeration of
all self-avoiding conformations for \(N \leq 15\) on  cubic
lattice.  In doing so we
calculated the energies of all conformations according to Eq. (19), and
then determined the number of MES.  Each quantity, such as the number
of MES, \(C\)(MES), the lowest energy \(E_{0}\), the number of
nearest-neighbor contacts \(t\) in the lowest energy structures, 
is averaged over 30 sequences.
Therefore, when referring to these quantities, we will imply their
average values.  To test
the reliability of the computational results an additional 
sample of 30 random
sequences was generated. Note that in the case of $C$(MES) 
we computed the quenched averages, i.e., 
\(C\text{(MES)}=exp[\overline{ln[c\text{(MES)}]}]\), where \(c\) is the
number of MES for one sequence.

The number of MES \(C\)(MES) is plotted as a function of the number of
residues \(N\) in Fig. (2) for \(B_{0}=-0.1\) and
\(\Delta=0.6\). A pair of squares at given \(N\) represents
\(C\)(MES) computed for two independent runs of 30 sequences each.
For comparison, the number of self-avoiding walks 
\(C\)(SAW) and the number of CS
\(C\)(CS) are also plotted in this figure (diamonds and triangles,
respectively). The most striking and important result of this graph is
the following:  As expected on general theoretical grounds, 
\(C\)(SAW) and \(C\)(CS) grow exponentially with
\(N\), whereas the number of MES \(C\)(MES) exhibits drastically
different  scaling
behavior. There is no variation in $C$(MES) and its value remains 
practically constant within the entire
interval of \(N\) starting with \(N=7\). We find (see Fig. (2)) that
$C(\text{MES})\approx 10^1$. This result further validates our earlier
finding for two dimensional model. These results suggest that 
$C$(MES) grows (in all likelihood) only as  $ln N$ with $N$.  Thus the
restriction of compactness and low energy of the native states may force
an upper bound on the number of distinct protein folds.

\noindent 
{\em 3D HP model:} The calculations described above suggest that upon
imposing minimal restrictions on the structures (compactness and low
energies) the structure space becomes sparse. As suggested before this
must imply that each basin of attraction (corresponding to a given
MES) in the structure space must contain numerous sequences. The way
these sequences are distributed among the very slowly growing number
(with respect to $N$) of MES, i.e., the density of sequences in
structure space, is an important question. This was beautifully 
addressed in the
paper by  Li
{\em et. al} \cite{Li96}. 
They considered a three dimensional ($N=27$) cubic
lattice. By using HP model and restricting themselves to only
maximally compact structures as putative native basins of attractions
(NBA) they showed certain basins have much larger number of
sequences. In particular, they discovered that one of the NBAs serves
as a ground state for 3794 (total number is $2^{27}$) sequences and
hence was considered most designable (Fig. (3)). The precise density of
sequences
among the NBAs is clearly a function of the interaction scheme. These
calculations and the arguments presented in the previous subsection using
the random bond model
point out that since
the number of NBA for the entire sequence space is small it is likely
that proteins could have evolved randomly.
Naturally occurring
folds must correspond to one of the basins of attraction in the structure
space so
that many sequences have these folds as the native conformations,
i.e., these are highly designable structures in the language of
LWC \cite{Li96}. 
These ideas have been further substantiated by Lindgard and Bohr 
\cite{Lindgard96},
who showed that among maximally compact structures there are only very
few folds that have protein-like characteristics. These authors also
estimated using geometrical characteristics  and stability arguments  
that the number of distinct
folds is on the order of a thousand. All of these studies confirm that
the  density of the structure
space is sparse. Thus, each fold can be designed by many
sequences. From the purely structural point of view nature does have several
options in the sense that many sequences can be "candidate proteins". 
However, there is also evolutionary pressure to fold rapidly 
(i.e., a kinetic component to folding). This requirement further
restricts the possible sequences that can be considered as
proteins, because  they must satisfy the dual criterion of reaching a
definite fold on a biologically  relevant time scale. These observations
are schematically sketched in Fig. (4).

\textit{Symmetry and designability:} In the study by
Li. et. al. \cite{Li96} it was noted that highly designable structures
appear to be
symmetric. Independently, in a thought
provoking article Wolynes \cite{Wolynes96} has 
made a series of compelling arguments as to why nature might use symmetry
(at least in an inexact manner) to generate symmetrical tertiary folds
of
proteins. Many enzymes are oligomers. Wolynes makes a number of
observations about the symmetry aspects of protein structures: (a) The
terameric hemoglobin ($\alpha $-helical protein)
molecule has an approximate two-fold symmetry. (b) A striking example of 
approximate symmetry in $\beta $ proteins is found in the 
the structure of a monomeric $\gamma $ crystallin in which the shapes
adopted by residues 1-88 and 89-174 are nearly the same. However, the two
individual sequences do not bear much similarity. This, of course, is
consistent with the notion that the structure space is so sparse that many
sequences are forced to adopt similar shapes. The interesting conclusion
from examining the $\gamma $ crystallin structure is that the underlying
symmetries in the shape are only inexact. (c) The obvious example of very
nearly symmetrical structures are in helical proteins with the the four
helix bundle being one the most prominent examples (see
Fig. (5)). (d) Various
proteins with mixed topology (like TIM barrels and jelly rolls) appear to
have the
kind
of inexact but apparent symmetrical arrangement discussed by Wolynes
\cite{Wolynes97}.
(e) He also conjectured that it is likely that the underlying
approximate symmetry is reflected in the free energy landscape being
funnel-like. This would facilitate rapid folding which for many proteins
may be a result of evolutionary pressure. The precise connections between
the symmetries and the folding mechanisms and functional competence of
biological molecules have not been worked out. Nevertheless it appears that
employing such ideas might be useful in \textit{de novo} design of
proteins.

We note that equally striking are the kind of symmetrical arrangements
found in RNA molecules \cite{Cate96}. The crystal structure of the P4-P6
domain of
\textit{Tetrahymena} self-splicing RNA clearly is highly symmetric with helices
packed  in a nearly regular arrangement. Since in an evolutionary sense the
RNA world might have preceded the protein world it is interesting to
speculate that the emergence of inexact design may have been a biological
necessity. The observation of inexact symmetries in a protein
structure might be a consequence of the fact that they are present in
the "parent" molecules. In fact this evolutionary conservation may
have been imprinted when evolution from RNA world to the current
scheme for protein synthesis took place. The most compelling reason for
observing near regular
patterns in biomolecular structures is 
because synthesis of symmetrical folds might be energetically economical.

\vspace{7mm}
\noindent
\textit {Exploring protein folding mechanism using lattice model:}  
It is well known that proteins reach the biologically
active native states in a relatively short time which is on the order of a
second for most single domain proteins \cite{Stryerbook}. Based on
folding and
refolding
experiments on ribonuclease A Anfinsen concluded that under appropriate
conditions natural sequences of proteins spontaneously fold to their native
conformation \cite{Anfinsen73}. This implies that protein 
folding is self-assembly process
i.e., the information needed for specifying the topology of the native state
is contained in the primary sequence itself. This thermodynamic hypothesis
does not, however, address the question of how the native state is accessed
in a short time scale. This issue was raised by Levinthal who wondered how a
polypeptide chain of reasonable length can navigate the astronomically large
conformational space so rapidly. 
Levinthal posited that certain preferred
pathways must guide the chain to the native state. The Levinthal paradox,
simplistic as it is, has served as an intellectual impetus to understand the
ease with which a polypeptide chain reaches the native conformation
\cite{Onuchic97,Dill97}.
We use lattice models to describe the foldability of biological
sequences of proteins. A sequence is foldable if it reaches the native state
in a reasonable time and remains stable over some range of external
conditions (pH, temperature).

\vspace{7mm}
\noindent

\vspace{7mm}

\noindent
\textit{Characteristic Temperatures:}  The basic features of folding
can be understood in terms of two fundamental equilibrium temperatures that
determine the "phases" of the system
\cite{Thirumalai99}. At sufficiently high temperatures
(\(kT\) greater than all the attractive interactions) the shape of the
polypeptide chain can be described as a random coil and hence its behavior
is the same as a self-avoiding walk. As the temperature is 
lowered one
expects a transition at \(T = T_{\theta }\) to a compact phase. This transition
is very much in the spirit of the collapse transition familiar in the theory
of homopolymers \cite{degennesbook}. The number of compact conformations
at \(T_{\theta }\)
is still exponentially large. Because the polypeptide chains have additional
energy scales that discriminate between the various compact conformations we
expect a transition to the ground (native) state at a lower
temperature 
\(T_{F}\). Generally the transition at \(T_{\theta }\) 
is second order, while the
transition at \(T_{F}\) is first order like. Such transitions here are for very
small systems and the notion of ''phases'' should be used with care. These
expectations, based on fairly general arguments, have been confirmed in
various lattice simulations of protein-like
heteropolymers. 
For the lattice models 
the collapse temperature \(T_{\theta } \) is determined from the peak of the
specific heat and the folding transition temperature is obtained from the
fluctuations in the overlap function given by
\begin{equation}
\Delta \chi =<\chi ^2>-<\chi >^2
\end{equation}
where
\begin{equation}
\chi =1 - \frac {1}{N^2-3N+2} \sum _{i<j+2} \delta 
(\vec{r_{ij}}-\vec{r}_{ij}^{N})
\end{equation}
with \(r_{ij}^N\) referring to the native state. 
In Fig. (7a) (for the structure displayed in Fig. (6)) we
plot the temperature dependence of \(C_{v}\), which has a peak at 
\(T_{\theta } = 0.83\).
This figure also shows the variation of \(d\,<\,R_{g}\,>\,/dT\) 
with temperature. The peak of this curve (\(0.86\)) almost 
coincides with that of the
specific heat indicating that this transition is associated with compaction
of the chain. Hence, the maximum in \(C_{v}\) 
legitimately indicates the  collapse temperature.  X-ray scattering
experiments have been used to obtain $T_\theta$ for a few proteins. 
In Fig. (7b) we show the temperature dependence of
\(\Delta
\chi \) from which the folding temperature \(T_{F}\) is determined to be 0.79.

\vspace{7mm}
\noindent
{\bf Folding Rates:} The key question we want to answer is 
what are the intrinsic sequence dependent factors that not only determine
the folding rates but also the stability of the native state? It turns
out that many of the global aspects
of folding kinetics of proteins  can be understood in terms of the
equilibrium transition temperatures. In particular, we will show that the key
factor that governs the foldability of sequences is the single
parameter 
\begin{equation}
\sigma _{T} = \frac{T_{\theta} -T_{F}}{T_{\theta } }
\label{sigma}
\end{equation}
which indicates how far \(T_{F}\) is from \(T_{\theta }\). 
To establish a
direct correlation between the folding time \(\tau _{F}\) and 
\(\sigma _{T}\) we
generated a number of sequences for \(N = 27\). 
The folding time was taken to be
equal to the mean first passage time. The first passage for a given initial
trajectory was calculated by determining the total number of Monte Carlo
steps (MCS) needed to reach the native conformation for the first time. By
averaging over an ensemble of initial trajectories (typically this number
varies between 400\,-\,800 in our examples) the mean first passage time is
obtained. The precise moves that are utilized in the simulations are
described elsewhere \cite{KlimovProt96}. The dependence of \(\tau
_{F}\) on 
\(\sigma _{T}\) (for the random bond model and for the other
interaction schemes)
is given in
Fig. (8). This figure shows a remarkable correlation between the folding
time and \(\sigma _{T}\). A small change in \(\sigma _{T}\) 
results in a dramatic
effect (a few orders of magnitude) on the folding times. It is clear that
both \(T_{\theta }\)  and \(T_{F}\) are dependent on the sequence. As a result
mutations that preserve the native state can alter the folding rates due to
the change in the \(\sigma _{T}\) values.

Using lattice models we have also established that folding rates
correlate well with $Z = (E_N-E_{MS}/\delta$, where $E_N$ is the
native state energy, $E_{MS}$ is the average energy of the ensemble of
misfolded structures, and $\delta$ is the dispersion in the contact
energies. The relationship between $\sigma $ and $Z$ also suggests
that, in general, the correlation between $\tau _F$ and $\sigma $
should be superior. More importantly, experimental measurements of $Z$
are difficult. On the other hand, both $T_\theta$ amd $T_F$ can be
measured in scattering, CD, or fluorescence experiments. Other
measures, such as energy gap (however, it is defined), do not correlate
with $\tau _F$. 

In the previous section we showed that because the structure space is
very sparse there have to be many sequences that map onto the
countable number of basins in the structure space. The kinetics here
shows that not all the sequences, even for highly designable
structures, are kinetically competent. Consequently, the biological
requirements of stability and speed of folding severely restrict the
number of evolved sequences for a given fold. This very important
result is schematically shown in Fig. (4). 

It is important to point out that the simulations reported in
Fig. (8) were
done at sequence dependent temperatures using the condition 
\(<\chi (T_{s})> = 0.21\). At these temperatures, all of which are
below their respective folding transition temperatures, the native
conformation has the highest occupation probability. In lattice models the
native state is a single conformation (a microstate) which is, of
course,  
physically unrealistic. In real systems there is a volume associated with
the native basin of attraction (NBA) and there are many conformations that
map onto the NBA. The probability of being in the NBA at the various simulation
temperatures is in excess of 0.5 so that under the conditions of our
simulations the {\em stability criterion is automatically satisfied}. 
The results
in Fig. (8), therefore, shows that the {\em 
dual requirement of stability and the
kinetic accessibility} of NBA is most easily satisfied by those sequences
that have small values of \(\sigma _{T}\). Thus rapid folding occurs when 
\(\sigma _{T} \approx  0\) i.e., near a multicritical-like point. In this case
there are no detectable 
intermediate "phases".The sequence, whose native state is
shown in Fig. (6) has \(\sigma _{T} = 0.05\). We found that this
sequence folds rapidly. 

\vspace{7mm}
\noindent
{\bf Topological Frustration and Kinetic Partitioning Mechanism:} Lattice
models can also be used to obtain the outlines of the mechanisms for folding
of proteins. The qualitative aspects of the folding kinetics of biomolecules
can be understood in terms of the concept of topological frustration. The
primary sequence of proteins has about 55\% hydrophobic residues. The linear
density of hydrophobic residues along the polypeptide chain is roughly
constant implying that the hydrophobic residues are spread throughout the
chain. As a result on any length scale \(l\) there is a propensity for the
hydrophobic residues to form tertiary contacts under folding conditions. The
resulting structures which are formed by contacts between residues that are
in proximity would be in conflict with the global fold corresponding to the
native state. The incompatibility of structures on local scales with the
near unique native state on the global scale leads to topological
frustration. It is important to realize that topological frustration is
inherent to all foldable sequences, and is a direct consequence of the
polymeric nature of proteins as well as competing interactions (hydrophobic
residues which prefer the formation of compact structures and hydrophilic
residues which are better accommodated by extended conformations). A 
consequence of topological frustration is that the underlying energy
landscape is rugged consisting of many minima that are separated by barriers
of varying heights. 

It is important to understand the nature of the low-lying minima in the
rugged energy landscape. On the length scale \(l\) there are many ways of
forming structures that are in conflict with the global fold. It is expected
that most of these structures have high free energies and are unstable to
thermal fluctuations. We expect a certain number of these structures to have
low free energies and be relatively deep minima.  
The overlap between these
structures and the native fold could be considerable and hence these
structures could be viewed as being native-like. These competing basins of
attraction (CBA) in which the polypeptide chain adopts native-like
structures can act as kinetic traps that will slow down the folding process. 

The basic consequences of topological frustration for mechanisms of folding
can be understood in terms of the kinetic partitioning mechanism (KPM)
\cite{Thirumalai99}.
Imagine an ensemble of denatured molecules in search of the native
conformation. This is the experimental situation that arises when the
concentration of denatured molecules is decreased. It is clear that
a fraction of molecules \(\Phi \) would reach the NBA rapidly without being
trapped in the low lying energy minima. The remaining fraction would be
trapped in the minima and only on longer time scales fluctuations enables
the chain to reach the NBA. The value of the partition factor \(\Phi \)
depends on the sequence and is explicitly determined by the
\(\sigma _{T}\)
value. Thus because of topological frustration the initial pool of
denatured molecules partitions into fast folders and slow folders that reach
the native state by indirect off-pathway processes. 

From the description of the kinetic   partitioning mechanism (KPM)
given above it follows that 
generically the time dependence of the fraction of molecules that have not
folded at time \(t\), \(P_{u}(t)\), is given by
\begin{equation}
P_{u}(t)=\Phi \exp (-\frac {t}{\tau _{NCNC}})+\sum_{k} 
a_{k}\exp (-\frac {t}{\tau _{k}})
\label{Pu}
\end{equation}
where \(\tau _{NCNC}\) is the time constant for reaching the native state by
the fast process, \(\tau _{k}\) is the time for escape from the CBA 
labeled \(k\),
and \(a_{k}\) is the ''volume'' associated with the \(k^{\text{th}}\) 
CBA. From this
consideration we expect that for a given sequence trajectories can be
grouped into those that reach the native conformation rapidly (\(\Phi\)  being
their fraction) and those that remain in one of the CBA for discernible
length of time. In Fig. (9a) we show an example of trajectory that
reaches
the native state directly from the random coil conformation. In contrast in
Fig. (9b) we show an example of a trajectory for the same sequence at the
same simulation temperature. This figure shows that on a very short time
scale the chain gets trapped in conformations other than the NBA and only on
long time scale it reaches the native state. This figure illustrates the
basic principle of KPM. If we perform an average over an ensemble of such
trajectories the kinetic result given in Eq. (\ref{Pu}) ensues. 

\vspace{7mm}
\noindent
{\bf Classifying Folding Mechanisms in terms of $\sigma _{T}$: } 
The various folding mechanisms expected in foldable  sequences may be
classified in terms of the \(\sigma _{T}\). 
We have already shown that sequences 
that fold extremely rapidly have very small values of \(\sigma _{T}\). Based on
our study of several model proteins as well as analysis of real proteins we
classify the folding kinetics of proteins as follows \cite{Thirumalai99}:

\vspace{7mm}
\noindent
{\bf Fast Folders: } For these sequences the value of \(\sigma _{T}\) is less
than a certain small value \(\sigma _{l}\). For such sequences the folding
occurs directly from the ensemble of unfolded states to the NBA. The free
energy surface is dominated by the NBA (or a funnel) and the volume
associated with NBA is very large. The partition factor $\Phi $ is near
unity so that these sequences reach the native state by two-state kinetics.
The amplitudes \(a_{k}\) in Eq. (\ref{Pu})  are nearly zero. 
There are no intermediates
in the pathways from the denatured state to the native state. Fast folders
reach the native state by a nucleation-collapse
mechanism which means that once a certain number of contacts (folding
nuclei) are formed then the native state is reached very rapidly 
\cite{Guo95,Shakh96}. The time
scale for reaching the native state for fast folders (which are normally
associated with those sequences for which topological frustration is
minimal) is found to be 
\begin{equation}
\tau _{NCNC} = \frac{\eta a}{\gamma } f(\sigma _{T})N^{\omega } 
\end{equation}
where $\eta $ is the solvent viscosity, $a$ 
is the typical size of a residue, 
$\gamma $ is the average surface tension between the residue and water, 
\(f(\sigma _{T})\) 
is typically an exponential function of \(\sigma _{T}\), and the exponent 
$\omega $ is between 3.8 and 4.2. In general, only small proteins ($N$
less than about 100) are fast folders.

\vspace{7mm}
\noindent
{\bf Moderate Folders: } Sequences for which \(\sigma _{l} 
\le \sigma _{T} \le \sigma _{h}\) 
can be classified as moderate folders. Unlike fast folding
sequences the $\Phi $ values are fractional which means that a substantial
fraction of molecules is essentially trapped in one of the CBAs for some
length of time. For these sequences there are detectable intermediates and
for all but very small proteins the rate determining step is the activated
transition from one of the CBAs to the native state. The average time scale
for transition from these misfolded structures to the native conformation is
given by 
\begin{equation}
\tau _{F} \approx \tau _{0}\exp (\sqrt{N})
\end{equation}
at \(T \approx T_{F}\). This shows that typical barriers for moderate folders
are quite small. As a result the folding times even for long proteins (\(N 
\approx  200\)) are only on the order of a second. It is these small barriers
that enable typical proteins to fold in biologically relevant time scale
without encountering the Levinthal paradox.

\vspace{7mm}
\noindent
{\bf Slow Folders and Chaperones:} For sequences with  \(\sigma _{T} \ge
\sigma _{h}\) folding is extremely slow and these sequences may not reach
the native state in a biologically relevant time scales. The volume
corresponding to NBA is very small in this case and as a result $\Phi $ is
nearly zero. The free energy surface is dominated by CBAs. Under these
circumstances spontaneous folding does not become viable. In cells such
proteins are rescued by chaperones. 
Typically this happens when \(N\) is so
large  that \(\tau _{0}\exp (\sqrt{N})\) exceeds reasonable folding time
scales. Thus in cells we expect that only those proteins which are large or
whose biological functioning state has to be oligomers require chaperones.

\textit{Minimum number of residues for obtaining foldable protein
structures:} Natural proteins are made up of twenty amino acid
residues. An important question, from the perspective of protein design,
is how many distinct types of residues are required for protein-like
behavior? Such a selection cannot be made arbitrarily because  in  natural
proteins one should have polar, hydrophobic, and charged residues. In
addition, for optimal packing of the core hydrophobic residues with
different van der Waals radii may be required.  
To explore the potential simplification of the
number of residues Wang and Wang \cite{Wang99} have carried out a highly
significant study using lattice models and standard statistical potentials
for the contact interaction elements $B_{ij} $ (Eq. (19)). They
discovered that a grouping of amino acid residues into five categories
mimics the folding behavior found using the standard twenty
residues. To demonstrate this they used a cubic lattice with $N = 27$,
and mostly focussed on the maximally compact structures as ground
states. Thus, structures such as ones given in Fig. (6) are not
explicitly considered. Nevertheless, the demonstration that a suitable
set of five amino acid residue types is sufficient is an important
result which should have implication for protein design problem - the
generation of primary sequences that can fold to a chosen target folded
structure.

In their original article they mostly focussed on various
thermodynamic properties (nature and degeneracy of the ground
states). They have also carried out kinetic simulations to assess if the
kinetic properties are altered by using a reduced number of residues. To
test this idea Wang and Wang used the foldability index $\sigma $ (which
correlates
well with fold rates) as a discriminator of sequence properties. The
precise question addressed by Wang and Wang is
the following: What is the minimum number of residues that are required to
obtain foldable (characterized by having relatively small values
of $\sigma $) sequences? We find that fast folding sequences have $\sigma
$ less than about a quarter.  They carried out two sets of
computations. In one set they initially optimized the stability gap
\cite{Onuchic97} of various sequences using the twenty residues. They
substituted the residues
in these optimized sequences by the representative residue for each group.
Four subgroups were considered with each
containing  five and a variant, three and two amino acids. The foldability
index
for the standard sample and their substitutes is shown in Fig. (10) as
solid circle. In another set of computations they examined the foldability
index (open diamonds in Fig. (10)) for sequences that were optimized using
the reduced sets of amino acids.  
Both these curves show that as long as the number of amino acid
types exceed five one can generate sequences with relatively small values
of $\sigma $. Fig. (10) also shows that smaller values of
$\sigma $ can be obtained if optimization is carried out with
reduced sets of amino acids. 
Such
sequences are foldable i.e, the dual requirements of stability over a wide
temperature range and the kinetic accessibility of their native states
are simultaneously satisfied.

\section{CONCLUSIONS}

The examples of modeling discussed in Sections (II) and (III) are meant
to
illustrate the ideas behind theoretical and computational approaches to
protein folding. It should be borne in mind that we have discussed only a
very limited aspect of the rich field of protein folding. The computations
described in Section (III) can be carried out easily on a desktop
computer. Such an exercise is, perhaps, the best of way of appreciating
the simple approach to get at the principles that govern the folding of
proteins. 

In this chapter we have not discussed experimental advances that are
offering extraordinary insights into the way the denatured molecules reach
the native state. Two remarkable experimental approaches hold the promise
that in short order we will be able to watch the folding process from
submicrosecond time scale till the native state is reached. A brief
summary of these follow.

1) Eaton and coworkers showed in 1993 that optical triggers of folding can
offer a window into the folding process from microsecond time scale
\cite{Eaton98}. Since
this pioneering work many laboratories have probed the plausible structure
formation that occur on short time scale. Fast folding
experimental techniques have been used to obtain 
detailed kinetics for the building blocks of proteins, namely,
$\beta $-hairpin, $\alpha $-helices, and loops. 
 Very recent experiments have
given compelling evidence that there are populated native-like
intermediates even in proteins that were thought to follow two-state
kinetics. 

2) Perhaps the most exciting development in the last few years is the
ability to nanomanipulate single biomolecules using atomic force
microscopy and optical tweezer techniques \cite{Fernandez99}. So far such
experiments have
been used to provide a microscopic basis of elasticity in muscle
proteins. If these stretching experiments can be combined with
fluorescent resonance energy transfer experiments then it is possible to
follow the folding of individual molecules as it passes through the
transition state to the native conformation. It has been suggested on
theoretical grounds that such two-dimensional single molecule experiments
can measure directly the distribution of folding rates (and the barrier
distribution) in much the same way mean first passage times are computed 
in
minimal protein models (see Section (III)). 

The challenges posed by these high precision experiments 
demand more refined models
and further developments in computational techniques. For the
theoretically inclined it will no longer be sufficient to describe
kinetics only in terms of energy landscapes. The wealth of data that are
being generated by experiments, such as the ones mentioned above, requires
quantitative understanding of the various factors that govern the pathways,
mechanisms, and the transition states in the folding process. These
challenging issues will make the area of biomolecular folding an engaging
one for many years to come.

\acknowledgements

We are grateful to John D. Weeks for useful comments and to 
Chao Tang for supplying Fig. (3). We are indebted to
Dr. J. Wang and Prof. W. Wang for kindly providing us with Figure
(10) prior to publication.

\appendix
\section{}

There are several versions of the random heteropolymer models. To keep the
discussions technically simple we will consider one case - the so called
random hydrophilic-hydrophobic chain whose phases  were studied by
Garel, Orland and Leibler (GLO) \cite{Garel94}. The GLO model consists of
a polymer chain
with $N$ monomers.
The GLO model can be viewed as a generalization of the
popular Edwards model which was introduced to understand swelling of
real homopolymer chains in good solvents\cite{degennesbook}. In the GLO
model the chain is
made up of
hydrophobic (hydrophilic)
residues that tend to collapse (swell) the chain when dispersed in a
solvent. The solvent mediated interactions at each site is assumed to be
random. The random interactions depend only on a given site $i$ and the
strength depends on the degree of hydrophilicity $\lambda _{i}$. Besides the
term accounting for chain connectivity there are two and higher body
interactions
that determine the shape of the chain. In the GLO model the two body
interaction
is
given by 
\begin{equation}
v_{ij}=v_{0}+\beta (\lambda _{i}+\lambda _{j})\delta [r_{i}-r_{j}]
\end{equation}
where $v_{0}$ is repulsive short range interaction, $\lambda _{i}$ is a
quenched random variable which is distributed as
\begin{equation}
P(\lambda _{i})=\frac{1}{\sqrt{2\pi \sigma ^{2}}}\exp [-\frac{(\lambda
_{i}-\lambda _{0})^{2}}{2\sigma ^{2}}].
\end{equation}
If the mean $\lambda _{0}$ is positive then the majority of the residues
are hydrophilic. Description of the collapsed phase of the chain requires
introducing three and and four-body interaction terms. Thus, the total
Hamiltonian is
\begin{equation}
\beta H=\frac{1}{2}\sum_{i\neq j}v_{ij}+\frac{1}{6}\sum_{i\neq j\neq
k}\omega _{3}\delta (r_{i}-r_{j})\delta (r_{i}-r_{k})+\frac{1}{24}%
\sum_{i\neq j\neq k\neq l}\omega _{4}\delta (r_{i}-r_{j})\delta
(r_{j}-r_{k})\delta (r_{k}-r_{l}).
\end{equation}
Since the charge variables $\lambda _{i}$ are quenched the thermodynamics of
the system requires averaging the free energy using the distribution $%
P(\lambda _{i})$ i.e.,
\begin{equation}
F=-k_{B}T\int \prod P(\lambda _{i})\ln Z(\lambda _{i})d\{\lambda _{i}\}.
\end{equation}
The average of $\ln Z(\lambda _{i})$ is most conveniently done using the
replicas through the relation
\begin{equation}
\ln Z=\stackrel{\lim }{n\rightarrow 0}\frac{Z^{n}-1}{n}.
\end{equation}
Using Eqs. (2-4) the required average can be carried out. This leads to a
complicated expression for $\overline{Z^{n}}$ where the bar indicates the
average over the quenched random variables $\lambda _{i}$. In terms of the
order parameters 
\begin{equation}
q_{ab}(r,r^{\prime }) = \int ds \delta(r_a(s)-r) \delta(r_b(s)-r')
\end{equation}
and
\begin{equation}
\rho _{a}(r)=\int ds \delta(r_a(s)-r)
\end{equation}
 ($a$ and $b$ are
replica indices) the expression for $\overline{Z^{n}}$ becomes
\begin{equation}
\overline{Z^{n}}=\int Dq_{ab}(r,r^{\prime })D\widehat{q}_{ab}(r,r^{\prime
})D\rho _{a}(r)D\phi _{a}(r)\exp [H_{eff}]
\end{equation}
where 
\begin{equation}
H_{eff}=G(q_{ab},\widehat{q}_{ab},\rho _{a},\phi _{a})+\ln \zeta (\widehat{q}%
_{ab},\phi _{a})
\end{equation}
with
\begin{eqnarray*}
G &=&\int dr\sum (i\rho _{a}\phi _{a}-(v_{0}+2\beta \lambda _{0})\frac{\rho
_{a}^{2}}{2}-\frac{\omega _{3}^{^{\prime }}}{6}\rho _{a}^{3}-\frac{\omega
_{4}}{24}\rho _{a}^{4}) \\
&&+\int dr\int dr^{\prime }\sum_{a<b}(iq_{ab}(r,r^{\prime })\widehat{q}%
_{ab}(r,r^{\prime })+\beta ^{2}\lambda ^{2}q_{ab}(r,r^{\prime })\rho
_{a}(r)\rho _{b}(r^{\prime }))
\end{eqnarray*}
and
\begin{equation}
\zeta (\widehat{q}_{ab},\phi _{a})=\int \prod_{a}Dr_{a}(s)\exp
(-H_{T}\{r_{a}(s)\}).
\end{equation}
In Eq. (10) $H_{T}\{r_{a}(s)\}$ is
\begin{equation}
H_T\{r_{a}(s)\}=\exp
[-\frac{d}{2a^{2}}\int_{0}^{N}ds\sum_{a}(\frac{dr}{ds}%
)^{2}-i\int_{0}^{N}ds\sum_{a}\phi _{a}(r_{a}(s))-i\int_{0}^{N}ds\sum_{a<b}%
\widehat{q}_{ab}(r_{a},r_{b})].
\end{equation}
and
\begin{equation}
\omega _{3}^{^{\prime }}=\omega _{3}-3\beta ^{2}\lambda ^{2}.
\end{equation}
The path integrals in Eq. (10) may be evaluated using the spectrum of the
effective n-body Hamiltonian
\begin{equation}
H_{n}=-\frac{d}{2a^{2}}\sum_{a}\nabla _{a}^{2}+\sum_{a}i\phi
(r_{a})+\sum_{a<b}i\widehat{q}_{ab}(r_{a},r_{b})
\end{equation}
in the limit of $n\rightarrow 0$. If $N$ is very large then we can use
ground state dominance to evaluate the spectrum of $H_{n}$. This gives
\begin{equation}
\varsigma (\widehat{q}_{ab},\phi _{a})\simeq \exp [-N\min_{\{\Psi
(r)\}}\{<\Psi \mid H_{n}\mid \Psi >-E_{0}(<\Psi \mid \Psi >-1\}]
\end{equation}
where $E_{0}$ is the ground state energy of $H_{n}$. GLO evaluated the
integral over $q_{ab}$ (Eq. (6)) by a saddle point approximation which
leads to 
\begin{equation}
i\widehat{q}_{ab}(r,r^{\prime })=-\beta ^{2}\lambda ^{2}\rho _{a}(r)\rho
_{b}(r^{\prime }).
\end{equation}
From the above equation it follows that in the mean-field limit replica
symmetry is not broken. This makes the GLO model conceptually simpler
to interpret than the random bond heteropolymer model discussed in the
Appendix. 

The total wavefunction $\Psi
\{r_{1},r_{2},.....r_{n}\}$ (Eq. (12)) is written as a product of single
particle
functions (Hartree approximation). The various integrals are evaluated in
the saddle point approximation. A simple Gaussian form for the trial one
particle wavefunction 
\begin{equation}
\phi (r)=(\frac{1}{2\pi R^{2}})^{\frac{d}{4}}\exp (-\frac{r^{2}}{2R^{2}})
\end{equation}
is chosen with $R$ being the single variational parameter. Upon performing
the Gaussian integrals the free energy per monomer $f$ becomes
\begin{equation}
\beta f=\frac{a^{2}}{8R^{2}}+\frac{1}{(2\sqrt{\pi })^{d}}\frac{(v_{0}+2\beta
\lambda _{0})}{2}\frac{N}{R^{d}}+\Omega 
\end{equation}
where
\begin{equation}
\Omega =(\frac{1}{(2\pi \sqrt{3})^{d}}\frac{\omega _{3}}{6}-(\frac{1}{2\pi }%
)^{d}(3^{-d/2}-2^{-d}))(\frac{N}{R^{d}})^{2}+(\frac{1}{(32\pi ^{3})})^{d/2}%
\frac{\omega _{4}}{24}(\frac{N}{R^{d}})^{3}.
\end{equation}
At low temperatures the shape of the chain is determined by the sign of
first term in Eq. (18). If the sign is negative then the positive four
body
term is required for a stable theory. 

The phase of the random hydrophobic-hydrophilic model is complicated
and depends on the value of $\lambda _{0} $ \cite{Garel94}. We only
describe the hydrophilic
case when $\lambda _{0}$ is positive. In this case there is a first order
transition to a collapsed state ($R\sim N^{-1/d}$) induced by the negative
three body term. GLO pointed out that this transition is neither the usual
$\theta $-point nor is  it a freezing temperature because there is no
replica
symmetry breaking. In fact, this collapse transition resembles that seen in
proteins where it is suspected that it is first order transition. The
microscopic origin of the first order transition upon  collapse of 
polypeptide chains is not
fully understood. Recent
arguments, suggest that it could arise because burial of hydrophobic
residues and accommodation of the hydrophilic ones at the surface of 
proteins 
in water requires some work and perhaps this assembly happens in a
discontinuous manner.

\section{}

In Section (II) we considered a variational-type theory to treat the
thermodynamics of the random hydrophobic-hydrophilic heteropolymer. Here we
describe a limiting behavior of the random bond model
\cite{Shakh89}. 
In this appendix we show that the random bond model in the the
compact phase is identical to the Random Energy Model (REM). Historically,
REM was proposed as caricature for proteins on phenomenological grounds 
\cite{Bryngelson87}. The
heteropolymer with random bond interactions was treated using a variational
theory which suggested that when the disorder increases beyond a limiting
value the chain undergoes a thermodynamic glass transition. The nature of
this transition is closely related to Potts glasses.

The random-bond heteropolymer is described by a Hamiltonian similar to Eq.
(1) except the short range two-body term $v_{ij}$ is taken to be random
with
a Gaussian distribution. In this case a three-body term with a positive
value of $\omega _{3}$ is needed to describe the collapsed phase.The
Hamiltonian is 
\begin{equation}
H=\sum_{i<j}(v_{0}+v_{ij})\delta (r_{i}-r_{j})+\sum_{i\neq j\neq k}\delta
(r_{i}-r_{j})\delta (r_{j}-r_{k}) 
\end{equation}
The distribution of the random couplings is given by 
\begin{equation}
P(v_{ij})=\frac{1}{\sqrt{2\pi v^{2}}}\exp (-\frac{v_{ij}^{2}}{2v^{2}}). 
\end{equation}

In the collapse phase the monomer density $\rho =\frac{N}{R^{3}}$ is
constant (for large $N$). Thus, the only conformation dependent term in Eq.
(A1) comes from the random two-body term. Because this term is linear
combination of Gaussian variables we expect that its distribution is also
Gaussian and hence, can be specified by the two moments. Let us calculate
the correlation $\overline{E_{1}E_{2}}$ between the energies $E_{1}$ and $%
E_{2}$ of two conformations \{$r_{1}^{(1)}\}$ and \{$r_{i}^{(2)}\}$ of the
chain in the collapsed state. The mean square of $E_{1}$ is 
\begin{equation}
\overline{E_{1}^{2}}=\frac{v^{2}}{2}\sum_{i,j}\delta
(r_{i}^{(1)}-r_{j}^{(1)})=\frac{v^{2}}{2}N\rho 
\end{equation}
which is independent of the collapsed conformation. Similarly, we have 
\begin{eqnarray*}
\overline{E_{1}E_{2}} &=&\frac{v^{2}}{2}\sum_{i,j}\delta
(r_{i}^{(1)}-r_{j}^{(1)})\delta (r_{i}^{(2)}-r_{j}^{(2)}) \\
&=&\frac{v^{2}}{2}\sum_{r,r^{\prime }}q_{12}^{2}(r,r^{\prime })
\end{eqnarray*}
where $q_{12}(r,r^{\prime })$ is the overlap between the two conformations.
Because 
\begin{equation}
\sum_{r,r^{\prime }}q_{12}(r,r^{\prime })=N
\end{equation}
and since the monomer density is constant we $q_{12}(r,r^{\prime })=\frac{%
\rho ^{2}}{N\text{.}}$ . This implies 
\begin{equation}
\overline{E_{1}E_{2}}=\frac{v^{2}}{2}\rho ^{2}.
\end{equation}
Thus the joint probability is 
\begin{equation}
\lim_{N\rightarrow \infty }\sim \exp (-\frac{E_{1}^{2}+E_{2}^{2}}{N\rho v^{2}%
})
\end{equation}
which is equivalent the behavior in uncorrelated REM. Thus, it is not a
surprise that in large dimensions (which are captured by variational type
treatments) the random bond heteropolymer model yields exactly the same
result as the REM.

\bibliographystyle{unsrt}

\clearpage

\centerline{\bf FIGURES}
\vspace{0.35cm}

\noindent
{\bf Fig. (1)} The 3D native structure of hemoglobin visualized using
RasMol 2.6 \cite{RasMol}. The linear sequence of
amino acids of hemoglobin is given below the figure.  

\noindent
{\bf Fig. (2)} Scaling of the number of MES $C$(MES) (squares) is
shown for  the hydrophobic parameter $B_0=-0.1$ and 
$\Delta =0.6$. Data are obtained 
for the cubic lattice. The pairs of squares for each $N$
represent the quenched averages for different samples of 30
sequences. The number of compact structures $C$(CS) and self-avoiding
conformations $C$(SAW) are also displayed to underscore the dramatic
difference of scaling behavior of  $C$(MES) and $C$(CS) (or
$C$(SAW)). It is clear that $C$(MES) remains practically 
flat, i.e. it grows no faster than $ln N$. 

\noindent
{\bf Fig. (3)}  Histogram of number of structures with a given
number of associated sequences $N_s$ for 3D 3x3x3 case, in a log-log
plot.

\noindent
{\bf Fig. (4)} Schematic illustration of the stages in 
the drastic reduction of
sequence space in the process of evolution to functionally competent
protein structures. 

\noindent
{\bf Fig. (5)} Native structure of 
acyl-coenzyme A binding protein (first NMR structure out of
29 deposited to PDB). The figure was created using
RasMol 2.6 \cite{RasMol}.

\noindent
{\bf Fig. (6)} 
The native conformation of fast folding 
sequence ($N=27$) with random
bond potentials is shown. This structure has \(c=22\) non-bonded
contacts, therefore it is not a maximally compact conformation for which 
$c = 28$. The figure was created using
RasMol 2.6 \cite{RasMol}. 

\noindent
{\bf Fig. (7)} (a) Thermodynamic functions  computed for the sequence
whose native state is shown in Fig. (3). (a) Specific heat \(C_{v}\)
(dotted line) and derivative of the radius of gyration with respect to
temperature \(dR_{g}/dT\) (dashed line) as a function of
temperature. The collapse temperature \(T_{\theta }\) 
is determined from the peak of
\(C_{v}\) and found to be 0.83. \(T_{\theta }\)  is very close to 
the temperature, 
at which \(dR_{g}/dT\) becomes maximum (0.86). This illustrates
that \(T_{\theta }\) is indeed associated  with the compaction of the
chain. Temperature dependence of fluctuations of overlap function
\(\Delta \chi \) is given by  solid line. 
The folding transition temperature \(T_{F}\) 
is obtained from the peak of \(\Delta \chi \) and for this sequence 
\(T_{F} = 0.79\). The curves are scaled to fit one plot. \\
(b) Time dependence of the fraction of unfolded molecules $P_u(t)$ for
the sequence 74 calculated at folding conditions $T_s \lesssim
T_F$. The function $P_u(t)$ is computed from a distribution of first
passage times $\tau_{1i} $. First passage time for a given initial
condition is the first time the trajectory reaches the native
conformation. Typically an adequately converged distribution is
obtained by averaging over several hundred initial conditions. For
the conditions used in this simulation 
folding is two-state, therefore, $P_u(t)$
is adequately fit with the single exponential (thick solid line). 
The folding time $\tau _F$ obtained from the fit is $1.4\times 10^6
MCS$.

\noindent
{\bf Fig. (8)} 
Plot of the folding times 
$\tau _{F}$ as a function of $\sigma _{T}$ for the 22 sequences. 
This figures shows that
under the external conditions when the NBA is the most populated there is a
remarkable correlation between $\tau _{F}$ and $\sigma _{T}$.  
The correlation
coefficient is 0.94. It is clear that over a four orders of magnitude
of folding times \(\tau _{F} \approx
\exp (-\sigma _{T}/\sigma _{0})\) where $\sigma _{0}$ is a
constant. In both panels the
filled and open circles are for the RB and KGS 27-mer models, 
respectively. The open squares are for $N = 36$.

\noindent
{\bf Fig. (9)} Examples of folding trajectories at \(T=T_{s}\) derived
from the condition \(<\,\chi (T_{s})\,> = 0.21\). 
(a) Fast folding trajectory as monitored by \(\chi
(t)\). It is seen that sequence reaches the native state very rapidly
in a two-state manner without being trapped in intermediates.  
The first passage time for this  trajectory is 277,912 MCS. 
(b) Slow folding trajectory for the same sequence. The sequence
becomes trapped in several intermediate states with large \(\chi \)
en route to the native state. The first passage time is 11,442,793
MCS. Notice that the time scales in both panels are  dramatically 
different.

\noindent
{\bf Fig. (10)} The figure gives the foldability $\sigma$ of 27-mer 
lattice chains with sets containing different number of amino
acids. The sets are generated according to scheme described in
\cite{Wang99}.
The set of 20 amino acids is taken as a standard sample. 
Each sequence with 20 amino acids is optimized to fulfill 
the stability gap \cite{Onuchic97}.  The residues in the  
standard samples are substituted with four different sets containing
smaller number of amino acids \cite{Wang99}.
The foldability of these substitution are 
indicated by the solid circles. The open diamonds 
correspond to  the sequences with same composition. However, the amino
acids are chosen from the reduced representation and the result sequence
is optimized using the stability gap \cite{Onuchic97}.

\newpage
\begin{center}

\begin{minipage}{10cm}
\[
\psfig{figure=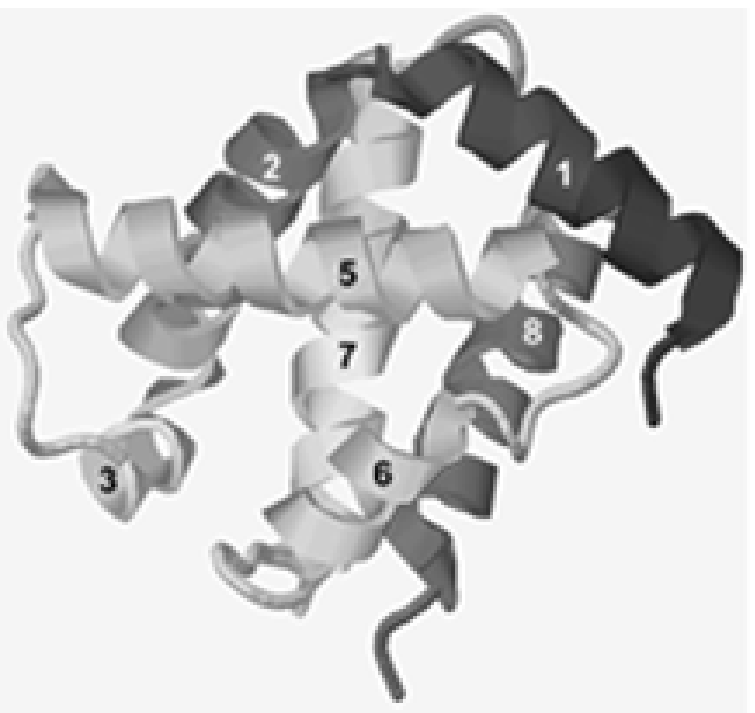,height=8cm,width=8cm}
\]
\end{minipage}
{\small \tightenlines

\begin{tabular}{ccccccccccccccccccccccc}

1val  &-& 2leu &-& 3ser &-& 4pro &-& 5ala &-& 6asp &-& 7lys &-& 8thr &-& 
9asn  &-& 10val &-& 11lys \\
12ala  &-& 13ala &-& 14trp &-& 15gly &-& 16lys &-& 17val &-& 18gly &-&
19ala &-& 20his &-& 21ala &-& 22gly\\ 
23glu &-& 24tyr &-& 25gly &-& 26ala  &-& 27glu &-& 28ala &-& 29leu &-&
30glu &-& 31arg &-& 32met &-& 33phe\\
34leu &-& 35ser &-& 36phe &-& 37pro &-& 38thr &-& 39thr &-& 40lys &-&
41thr &-& 42tyr &-& 43phe &-& 44pro\\
45his &-& 46phe &-& 47asp &-& 48leu &-& 49ser &-& 50his &-& 51gly &-&
52ser &-& 53ala &-& 54gln &-& 55val\\ 
56lys &-& 57gly &-& 58his &-& 59gly &-& 60lys &-& 61lys &-& 62val &-&
63ala &-& 64asp &-& 65ala &-& 66leu\\
67thr &-& 68asn &-& 69ala &-& 70val &-& 71ala &-& 72his &-& 73val &-&
74asp &-& 75asp &-& 76met &-& 77pro\\ 
78asn &-& 79ala &-& 80leu &-& 81ser &-& 82ala &-& 83leu &-& 84ser &-& 
85asp &-& 86leu &-& 87his &-& 88ala\\
89his &-& 90lys &-& 91leu &-& 92arg &-& 93val &-& 94asp &-& 95pro &-&
96val &-& 97asn &-& 98phe &-& 99lys\\ 
100leu &-& 101leu &-& 102ser &-& 103his &-& 104cys &-& 105leu &-&
106leu &-& 107val &-& 108thr &-& 109leu &-& 110ala\\ 
111ala &-& 112his &-& 113leu &-& 114pro &-& 115ala &-& 116glu &-&
117phe &-& 118thr &-& 119pro &-& 120ala &-& 121val\\ 
122his &-& 123ala &-& 124ser &-& 125leu &-& 126asp &-& 127lys &-&
128phe &-& 129leu &-& 130ala &-& 131ser &-& 132val\\ 
133ser &-& 134thr &-& 135val &-& 136leu &-& 137thr &-& 138ser &-& 
139lys &-& 140tyr &-& 141arg & &        & &        & &  

\end{tabular}
}

\vspace{10pt}

{\bf Fig. (1)}
\end{center}
\newpage

\newpage

\begin{center}
 
\begin{minipage}{15cm}
\[
\psfig{figure=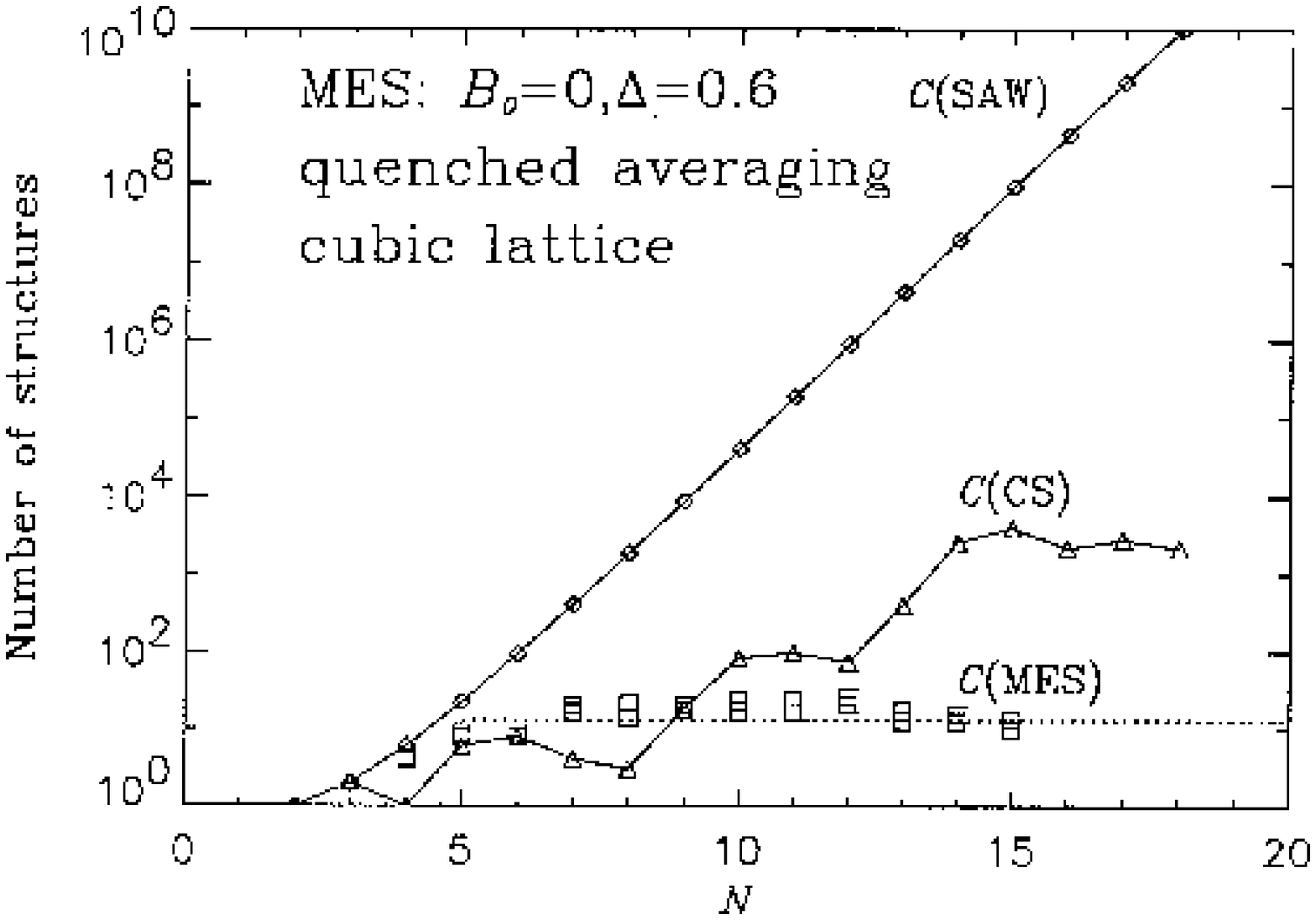,height=9cm,width=12cm}
\]
\end{minipage}

\vspace{5cm}

{\bf Fig. (2)} 
\end{center}

\newpage

\begin{center}

\begin{minipage}{15cm}
\[
\psfig{figure=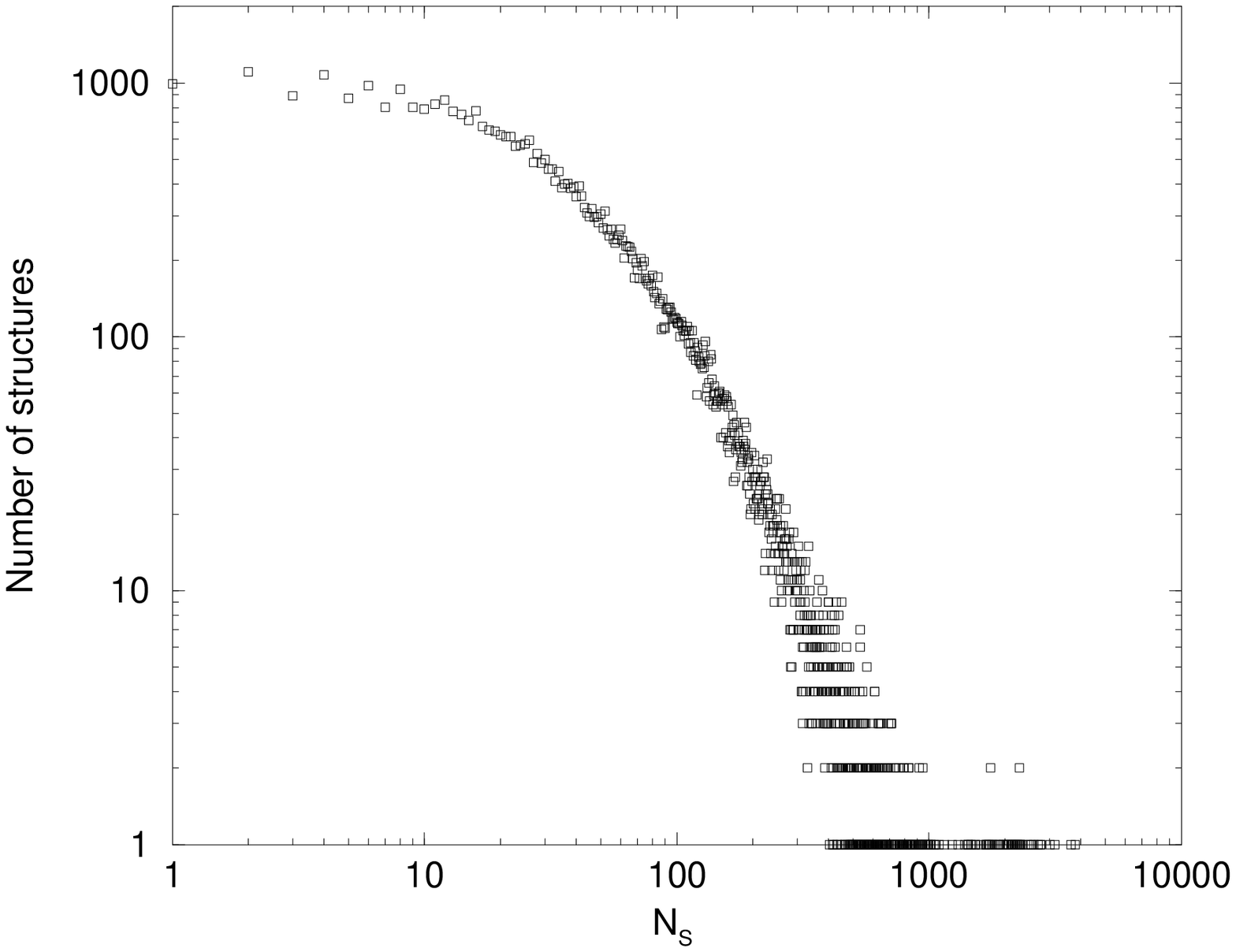,height=9cm,width=12cm}
\]
\end{minipage}

\vspace{0.5in}

{\bf Fig. (3)} 
\end{center}

\newpage

\begin{center}

\begin{minipage}{15cm}
\[
\psfig{figure=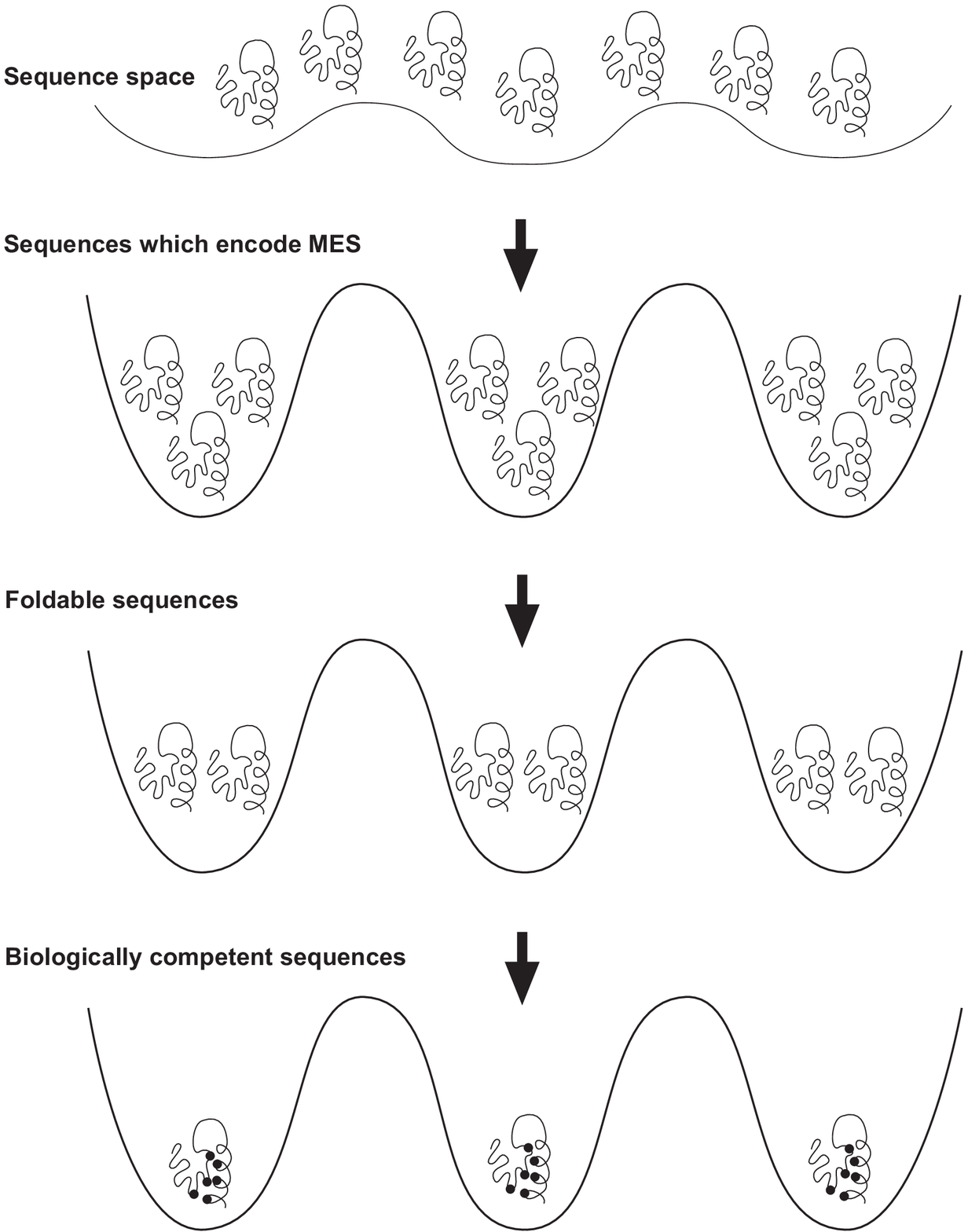,height=20cm,width=16cm}
\]
\end{minipage}

\vspace{0.5in}

{\bf Fig. (4)} 
\end{center}

\newpage

\begin{center}

\begin{minipage}{10cm}
\[\psfig{figure=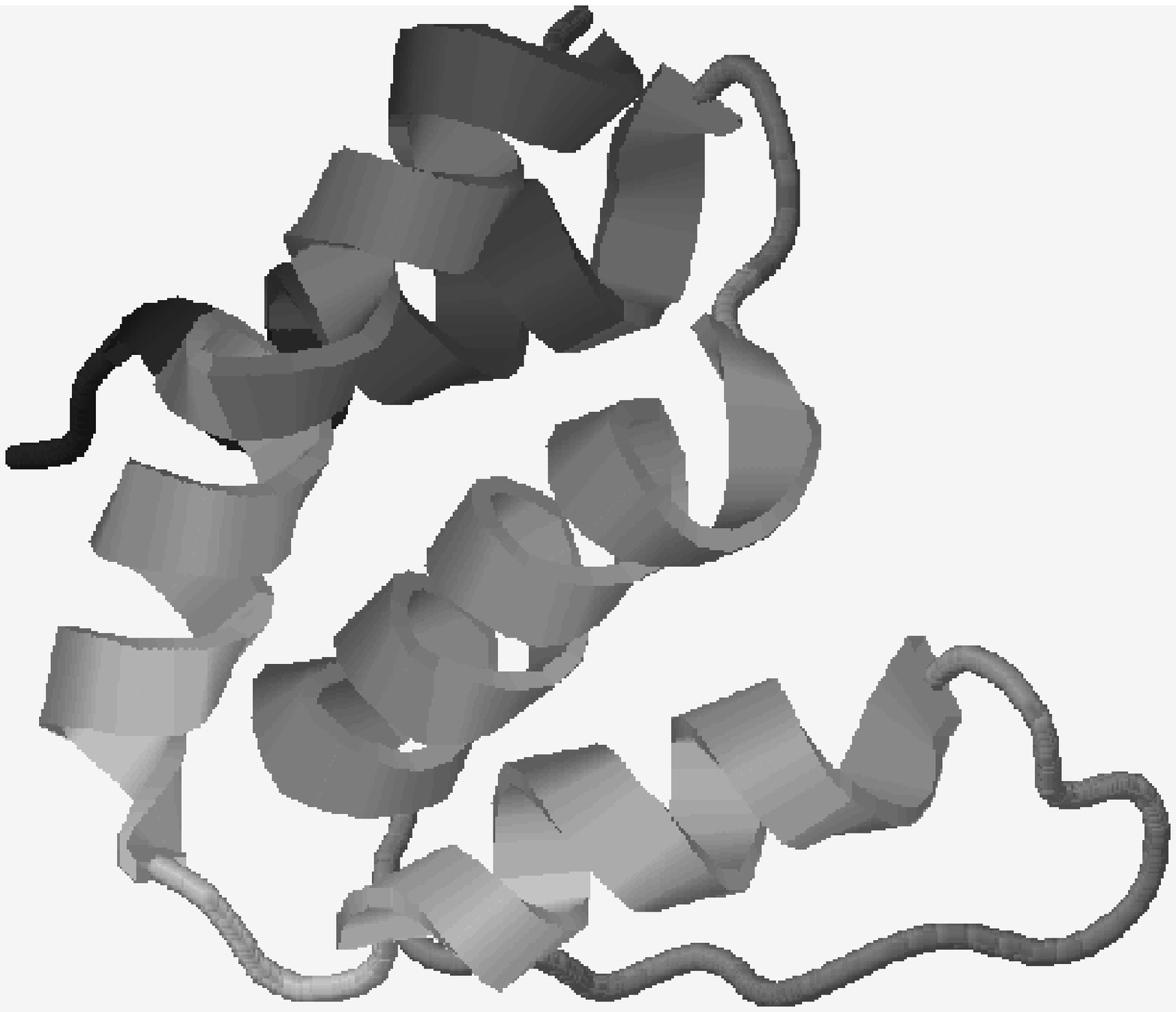,height=10cm,width=10cm,angle=-90}
\]
\end{minipage}

{\bf Fig. (5)}
\end{center}

\newpage

\begin{center}
\begin{minipage}{10cm} 
\[
\psfig{figure=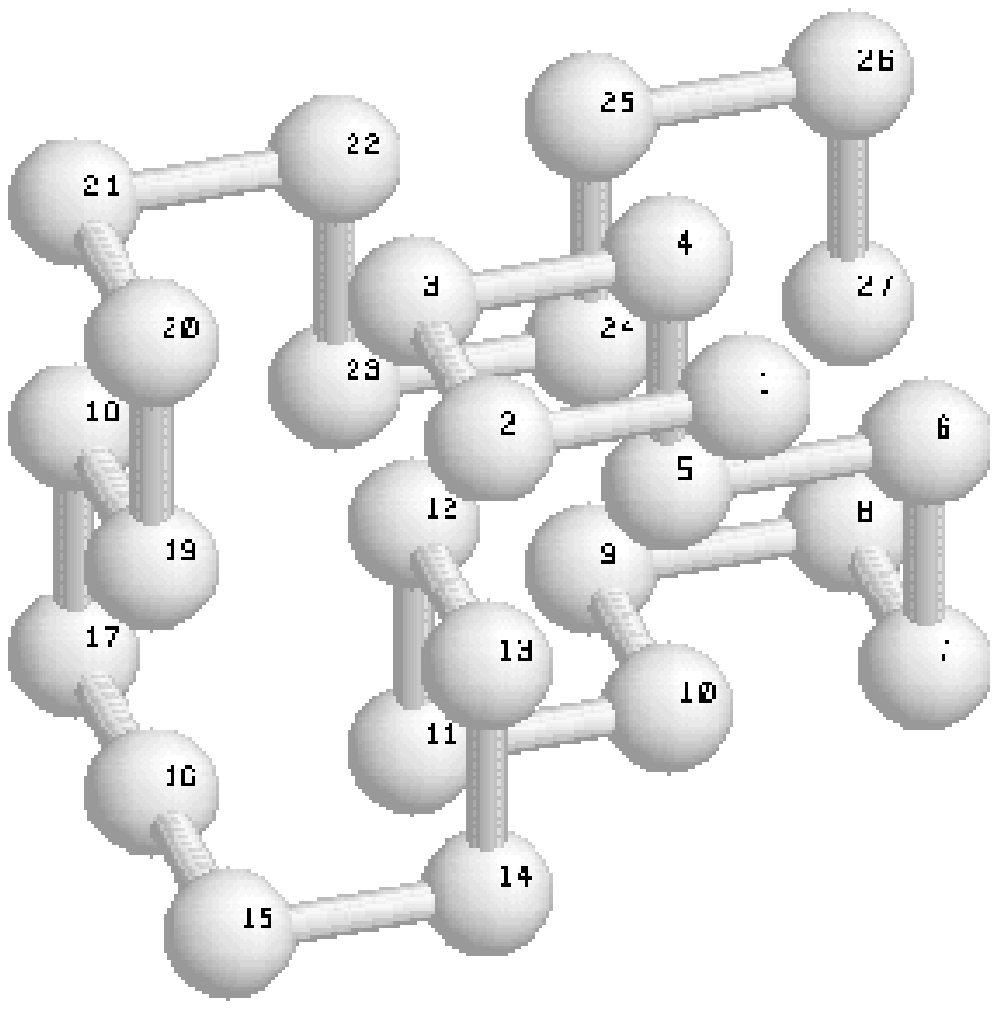,height=10cm,width=10cm}
\]
\end{minipage}

{\bf Fig. (6)} 

\end{center}
\newpage

\begin{center}
\begin{minipage}{15cm}
\[
\psfig{figure=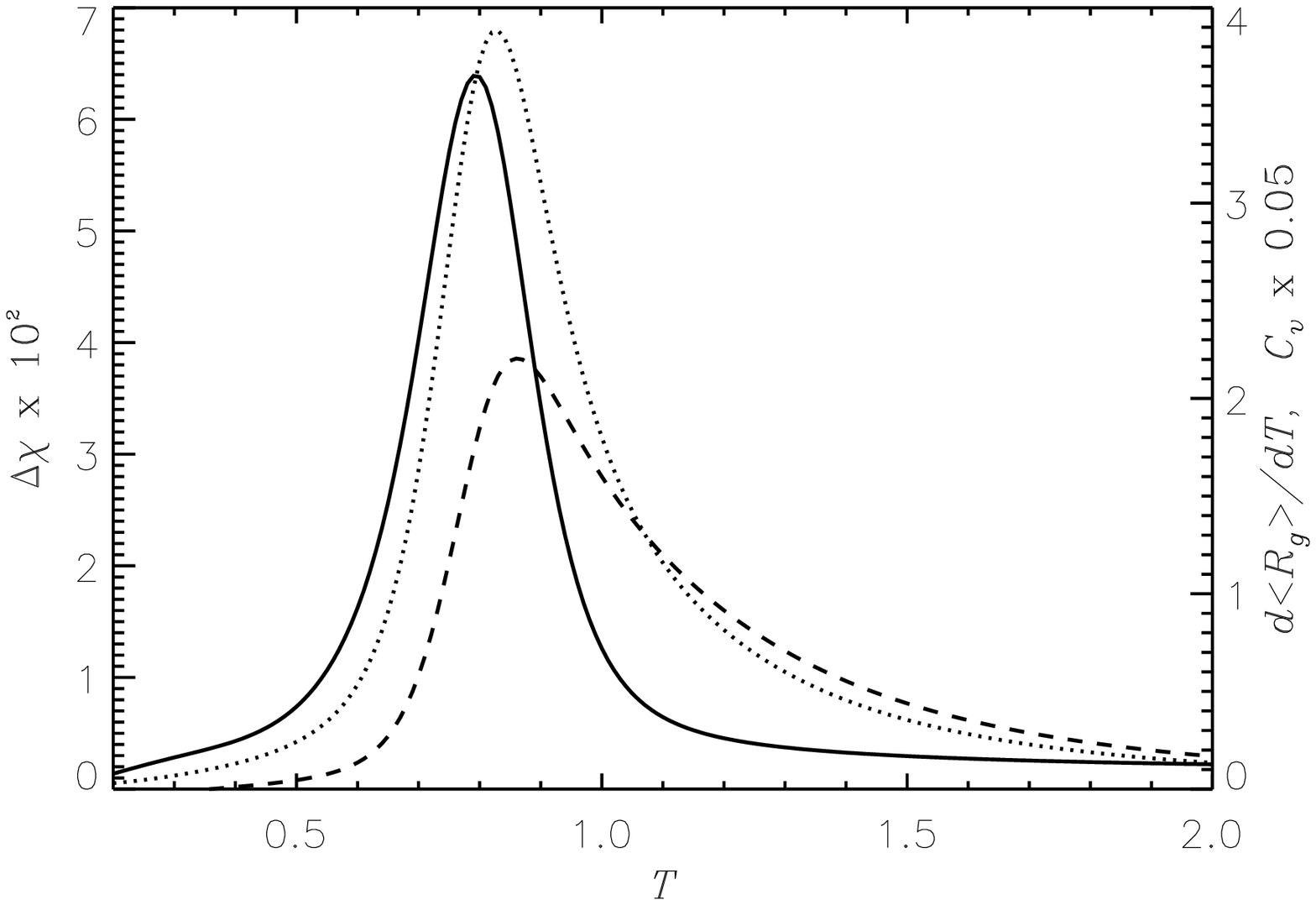,height=9cm,width=12cm}
\]
\end{minipage}

\end{center}

\begin{center}
\begin{minipage}{15cm}
\[
\psfig{figure=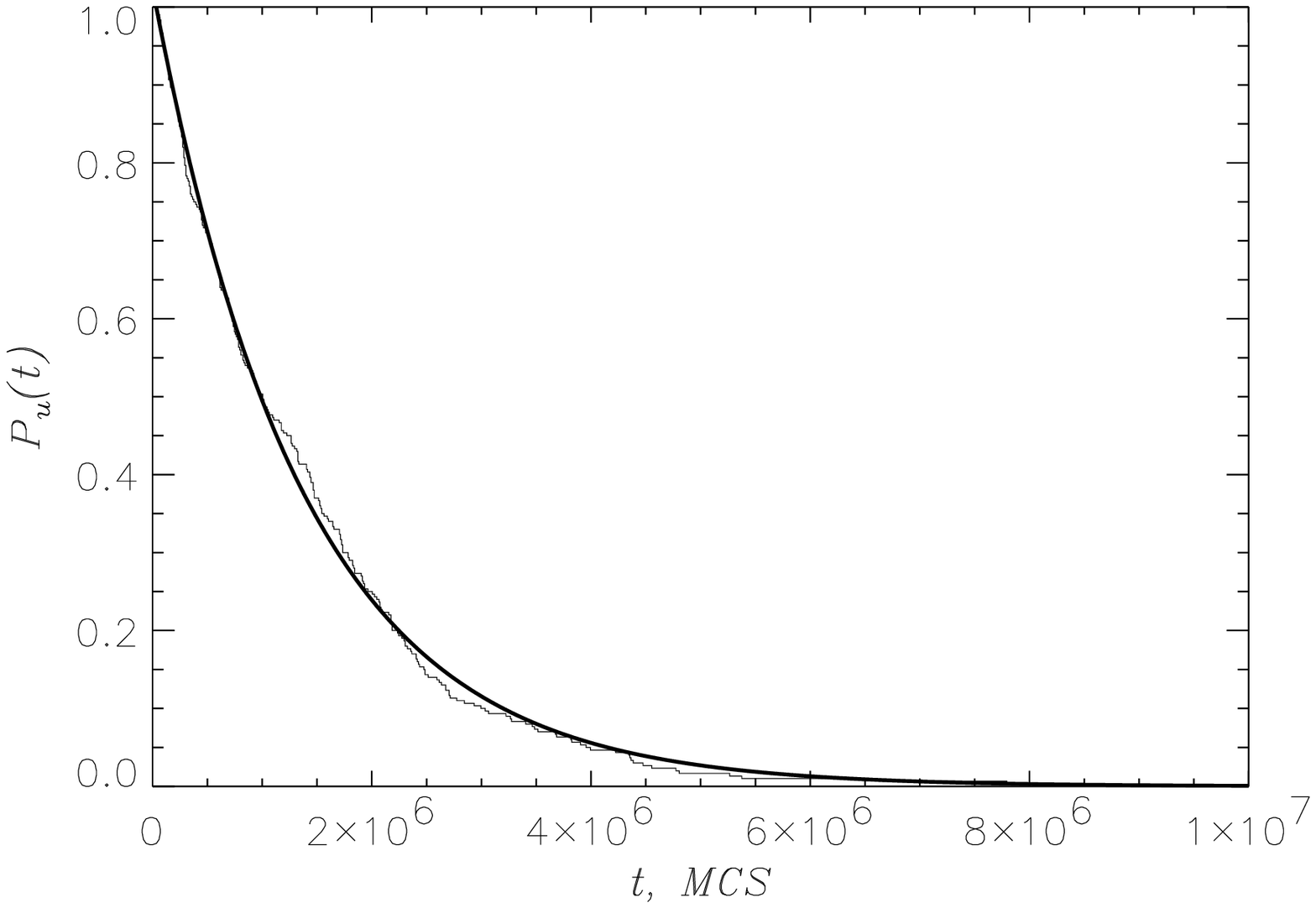,height=9cm,width=12cm}
\]
\end{minipage}

{\bf Fig. (7)} 
\end{center}

\newpage

\begin{center}
\begin{minipage}{15cm}
\[
\psfig{figure=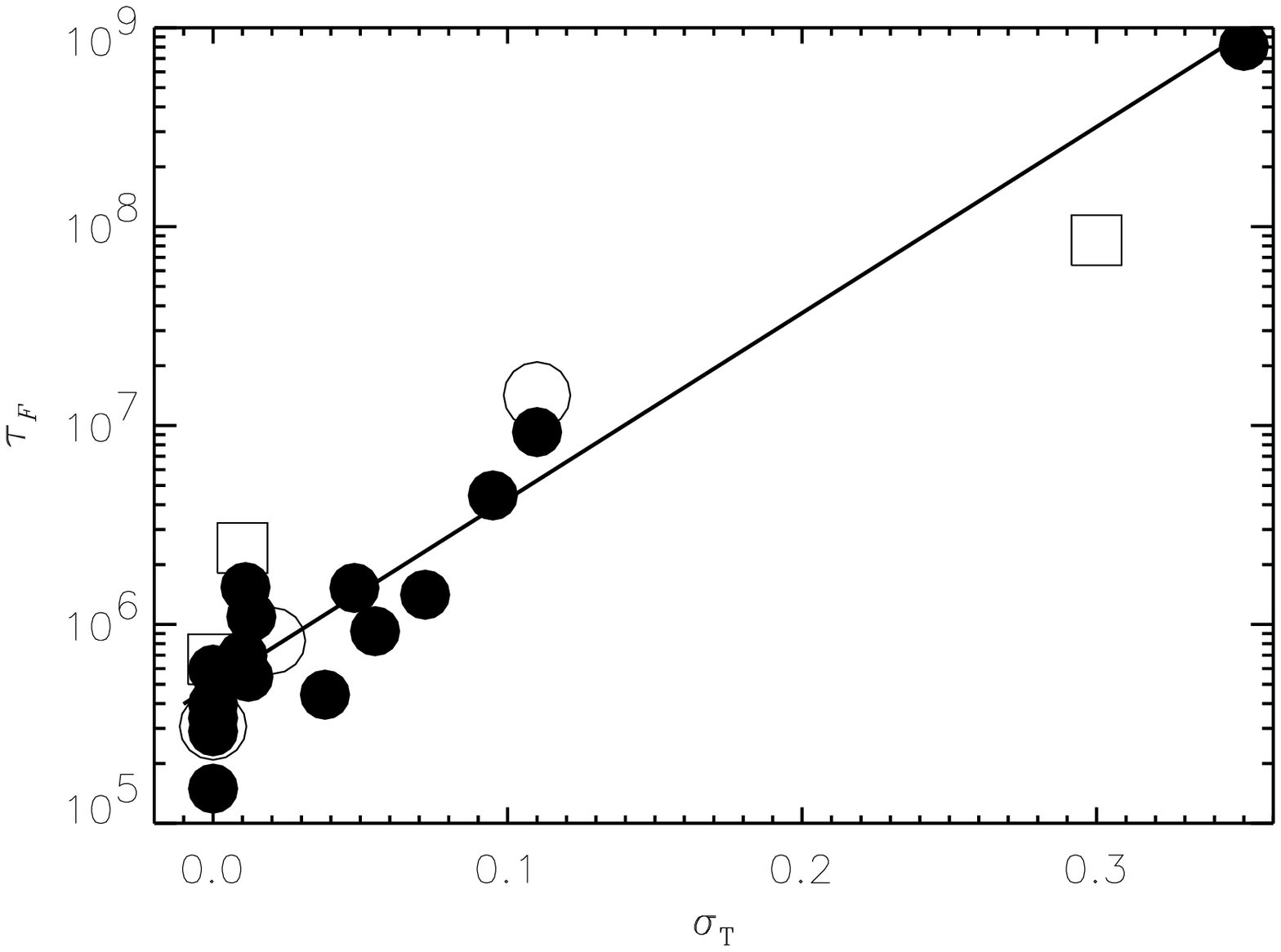,height=9cm,width=12cm}
\]
\end{minipage}

{\bf Fig. (8)} 

\end{center}

\newpage

\begin{center}
\begin{minipage}{12cm}
\[
\psfig{figure=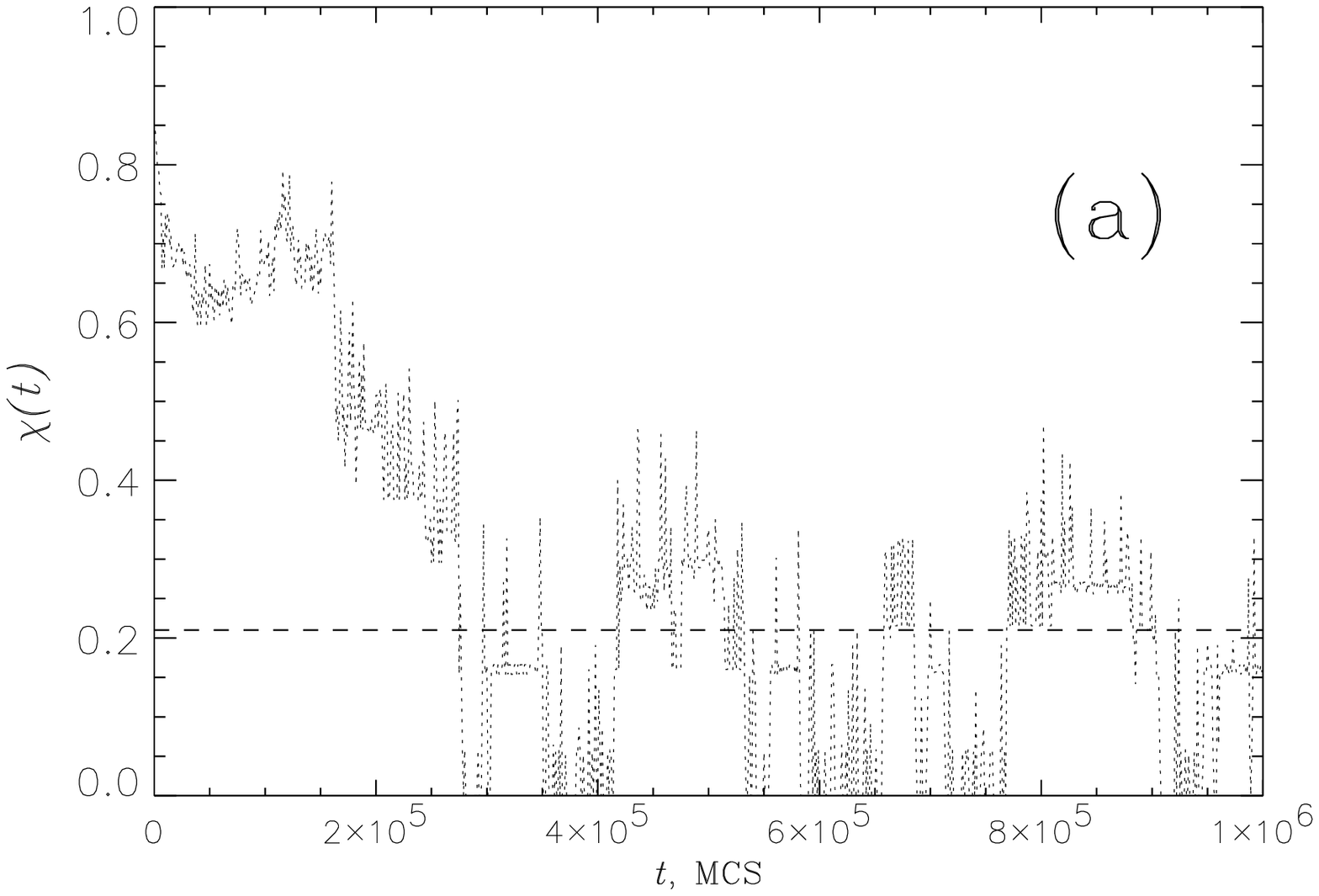,height=9cm,width=12cm}
\]
\end{minipage}
\end{center}

\begin{center}
\begin{minipage}{12cm}
\[
\psfig{figure=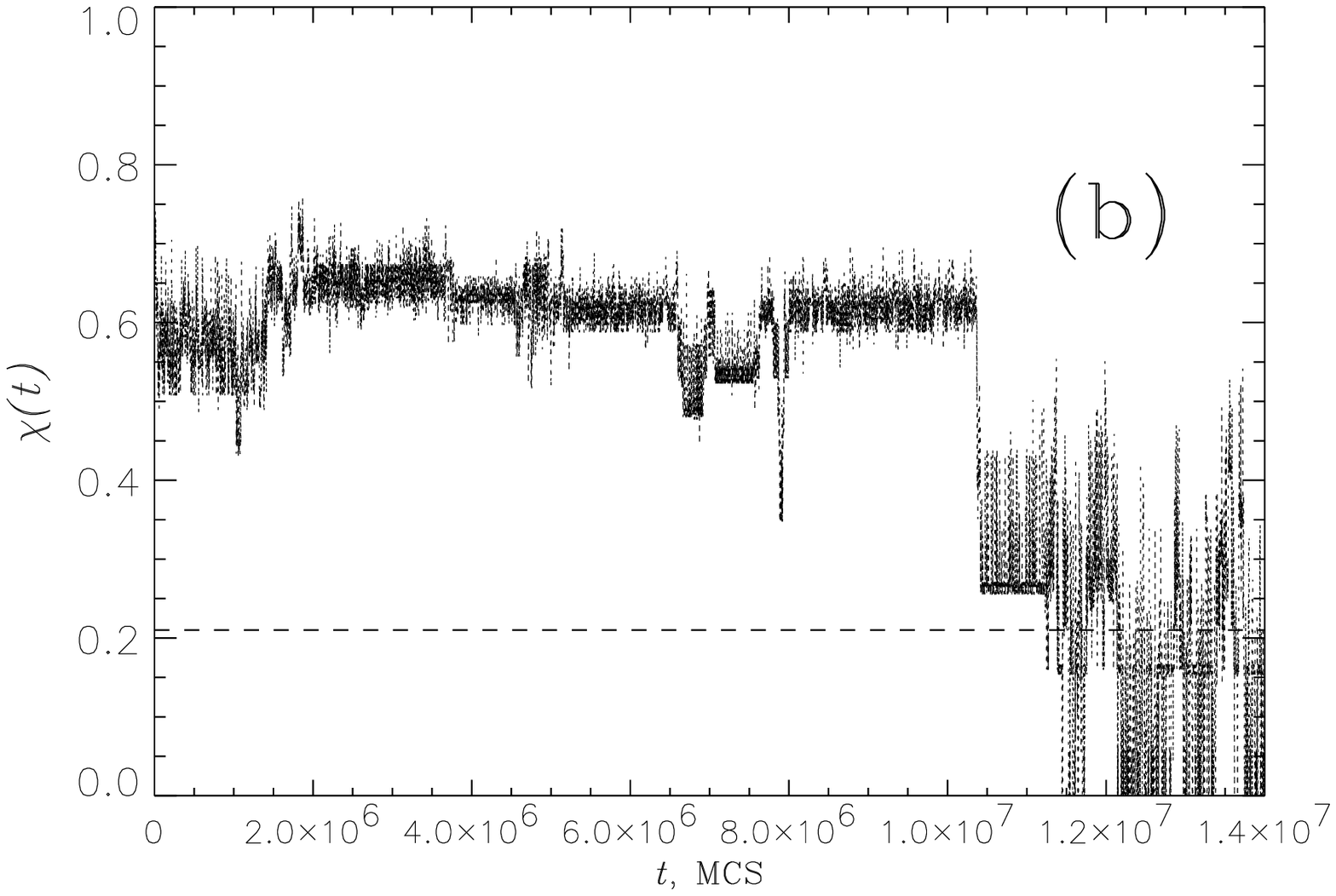,height=9cm,width=12cm}
\]
\end{minipage}

{\bf Fig. (9) }  

\end{center}

\newpage

\begin{center}
\begin{minipage}{12cm}
\[
\psfig{figure=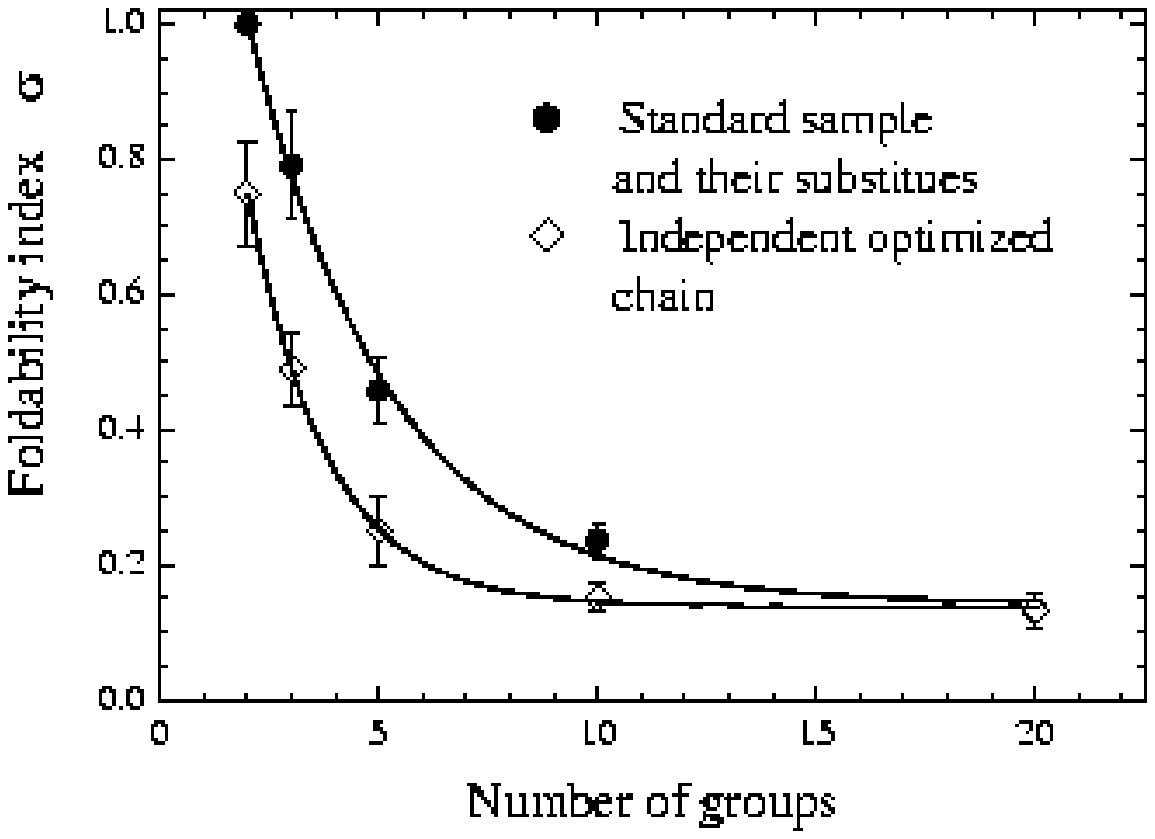,height=9cm,width=12cm}
\]
\end{minipage}

\vspace{30pt}

{\bf Fig. (10) }  

\end{center}

\end{document}